\documentclass[english]{article}
\usepackage[T1]{fontenc}
\usepackage[latin9]{inputenc}
\usepackage{color}
\usepackage{babel}
\usepackage{mathrsfs}
\usepackage{amsmath}
\usepackage{amssymb}
\usepackage{graphicx}
\usepackage{setspace}
\usepackage{esint}
\usepackage[unicode=true,pdfusetitle,
 bookmarks=true,bookmarksnumbered=false,bookmarksopen=false,
 breaklinks=false,pdfborder={0 0 1},backref=false,colorlinks=false]
 {hyperref}

\makeatletter

\providecommand{\tabularnewline}{\\}

\usepackage{cite}

\pdfoutput=1

\makeatother

\begin{document}
\noindent \begin{flushleft}
\textbf{\Large{}Optimized brute-force algorithms for the bifurcation
analysis of a spin-glass-like neural network model}\\

\par\end{flushleft}{\Large \par}

\noindent \begin{flushleft}
Diego Fasoli$^{1,2,\ast}$, Stefano Panzeri$^{1}$ 
\par\end{flushleft}

\medskip{}

\noindent \begin{flushleft}
\textbf{{1} Laboratory of Neural Computation, Center for Neuroscience
and Cognitive Systems @UniTn, Istituto Italiano di Tecnologia, 38068
Rovereto, Italy}
\par\end{flushleft}

\noindent \begin{flushleft}
\textbf{{2} Center for Brain and Cognition, Computational Neuroscience
Group, Universitat Pompeu Fabra, 08002 Barcelona, Spain}
\par\end{flushleft}

\noindent \begin{flushleft}
\textbf{$\ast$ Corresponding Author. E-mail: diego.fasoli@upf.edu}
\par\end{flushleft}

\section*{Abstract}

\noindent Bifurcation theory is a powerful tool for studying how the
dynamics of a neural network model depends on its underlying neurophysiological
parameters. However, bifurcation theory has been developed mostly
for smooth dynamical systems and for continuous-time non-smooth models,
which prevents us from understanding the changes of dynamics in some
widely used classes of artificial neural network models. This article
is an attempt to fill this gap, through the introduction of algorithms
that perform a semi-analytical bifurcation analysis of a spin-glass-like
neural network model with binary firing rates and discrete-time evolution.
Our approach is based on a numerical brute-force search of the stationary
and oscillatory solutions of the spin-glass model, from which we derive
analytical expressions of its bifurcation structure by means of the
state-to-state transition probability matrix. The algorithms determine
how the network parameters affect the degree of multistability, the
emergence and the period of the neural oscillations, and the formation
of symmetry-breaking in the neural populations. While this technique
can be applied to networks with arbitrary (generally asymmetric) connectivity
matrices, in particular we introduce a highly efficient algorithm
for the bifurcation analysis of sparse networks. We also provide some
examples of the obtained bifurcation diagrams and a Python implementation
of the algorithms.

\section{Introduction \label{sec:Introduction}}

\noindent Neural complexity refers to the wide variety of dynamical
behaviors that occur in neural networks \cite{Cessac1995,Pasemann2002,Fasoli2016}.
This set of dynamical behaviors includes variations in the number
of stable solutions of neuronal activity, the formation of neural
oscillations, spontaneous symmetry-breaking, chaos and much more \cite{Izhikevich2007,Ashwin2016}.
Qualitative changes of neuronal activity, also known as \textit{bifurcations},
are elicited by variations of the network parameters, such as the
strength of the external input to the network, the strength of the
synaptic connections between neurons, or other network characteristics.

Bifurcation theory is a standard mathematical formalism for studying
neural complexity \cite{Kuznetsov1998}. It allows the construction
of a map of neuronal activity, known as \textit{bifurcation diagram},
that links points or sets in the parameters space to their corresponding
network dynamics. In the study of firing-rate network models, bifurcation
theory has been applied mostly to graded (smooth) neural networks
with analog firing rates (e.g. \cite{Borisyuk1992,Beer1995,Cessac1995,Pasemann2002,Haschke2005,Fasoli2016}),
proving itself as an effective tool for deepening our understanding
of network dynamics. The bifurcation analysis of smooth models is
based on differential analysis, in particular on the Jacobian matrix
of the system. However, the Jacobian matrix is not defined everywhere
for artificial neuronal models with discontinuous activation function,
such as networks of binary neurons. For this reason, bifurcation theory
of smooth dynamical systems cannot be applied directly to these models.
On the other hand, while the bifurcation analysis of non-smooth dynamical
systems has recently received increased attention, it has been developed
only for continuous-time models, described by non-smooth differential
equations or by differential inclusions \cite{Leine2000,Awrejcewicz2003,Leine2006,Makarenkov2012}
(see \cite{Harris2015} for an example of application to firing-rate
models). Yet, many interesting problems in neuroscience involve the
use of non-smooth discrete-time models \cite{Karayiannis1993}, therefore
this gap prevents us from understanding the changes of dynamics in
these artificial systems.

If we define the states of a binary network of size $N$ as the collection
of the rates at which its cells are firing, then the number of possible
\textit{stationary} states is $2^{N}$. Moreover, the number of possible
\textit{oscillatory} solutions is $\sim\left(2^{N}-1\right)!$ for
large $N$, as we prove in Appx.~(\hyperref[sec:Appendix-A]{A}).
In particular, the bifurcation analysis of the model would unveil
the actual solutions in these large sets of states, for any given
set of network parameters. This would represent a tremendous simplification
in our comprehension of neural dynamics, providing great insight into
the operation of the network \cite{Little1974}.

In this article, we develop brute-force algorithms for the bifurcation
analysis of the spin-glass-like neural network model introduced in
\cite{Fasoli2017}. The network is composed of an arbitrary number
of neurons with binary firing rates, connected through arbitrary (generally
asymmetric) synaptic connections. The network dynamics is deterministic
and evolves in discrete time steps. In \cite{Fasoli2017} we proved
that the bifurcation structure of the model can be studied semi-analytically,
and in particular we were able to characterize the formation of multistability,
neural oscillations and symmetry breaking in the stimuli space. While
in \cite{Fasoli2017} we derived the bifurcation diagrams of simple
networks through hand calculations, in the present article we propose
algorithms for automatic bifurcation analysis. Therefore our work
complements standard numerical continuation softwares that are widely
used for the bifurcation analysis of graded neuronal models, such
as the MatCont Matlab toolbox \cite{Dhooge2003} and XPPAUT \cite{Ermentrout2002}.

Previous work focused on the emergence of complexity in ideal mean-field
limits of spin-glass models, see e.g. \cite{Sherrington1976,deAlmeida1978,Mezard1984}.
On the contrary, in the present article we do not make use of any
mean-field approximation. In particular, we consider exactly solvable
finite-size networks to be used in real-world applications. In Sec.\textcolor{blue}{~}(\ref{sec:Materials-and-Methods})
we briefly review the model (SubSec.\textcolor{blue}{~}(\ref{sub:The-Network-Model}))
and we show how the analytical formula of the state-to-state transition
probability matrix can be inverted for any network size $N$ to determine
the portions of parameters space where multistability, neural oscillations
and symmetry breaking occur (SubSec.\textcolor{blue}{~}(\ref{sub:Plotting-the-Bifurcation-Diagrams})).
The resulting formula can be written in any programming language.
In particular, we propose a Python implementation, that the reader
may find in the Supplementary Materials (scripts ``Multistability\_Diagram.py''
and ``Oscillation\_Diagram.py'').

However, the inversion of the state-to-state transition probability
matrix requires a priori knowledge of sets of candidate stationary
and oscillatory solutions, as we showed in \cite{Fasoli2017}, and
as we will explain in more detail in SubSec.\textcolor{blue}{~}(\ref{sub:Plotting-the-Bifurcation-Diagrams}).
For this reason, in SubSec.\textcolor{blue}{~}(\ref{sub:Non-Efficient-Algorithms-for-Generating-the-Stationary-and-Oscillatory-Solutions})
we introduce non-optimized algorithms that, through a brute-force
searching procedure, generate the sets of all the potential stationary
and oscillatory solutions (see the supplemental script ``Non\_Efficient\_Algorithm.py''
for a Python implementation). These algorithms can be applied to networks
with any topology of the synaptic connections (dense or sparse), but
their computational time increases exponentially with the network
size.

Yet, the study of sparse systems is of particular interest in neuroscience.
The average density of the synaptic connections, defined as the ratio
between the actual and the maximum possible number of connections
in the network, is approximately $10^{-6}-10^{-7}$ across the whole
cortex, and it can increase up to $\sim0.2-0.4$ in connection pathways
linking cortical areas \cite{Kotter2003}. For this reason, in SubSec.\textcolor{blue}{~}(\ref{sub:An-Efficient-Algorithm-for-Generating-the-Stationary-and-Oscillatory-Solutions-in-Sparse-Networks})
we propose an optimized brute-force algorithm, whose efficiency in
generating sets of candidate stationary and oscillatory solutions
increases with the network sparseness (see the supplemental script
``Sparse\_Efficient\_Algorithm.py'' for a Python implementation).

Then, in SubSec.\textcolor{blue}{~}(\ref{sub:Examples-of-Network-Topologies})
we introduce some widely-used examples of network topologies to be
tested, while in Sec.\textcolor{blue}{~}(\ref{sec:Results}) we show
the corresponding bifurcation diagrams generated by our codes. To
conclude, in Sec\textcolor{blue}{.~}(\ref{sec:Discussion}) we discuss
the importance and the biological implications of our results. In
particular, we discuss how our work advances the comprehension of
neural networks with respect to previous work (SubSec.\textcolor{blue}{~}(\ref{sub:Progress-with-Respect-to-Previous-Work-on-Bifurcation-Analysis})),
new insights into the dynamics of binary network models revealed by
our algorithms (SubSec.\textcolor{blue}{~}(\ref{sub:New-Insights-into-the-Dynamics-of-Discrete-Network-Models})),
and future directions that need to be pursued to address the limitations
of our work (SubSec.~(\ref{sub:Future-Directions})).

\section{Materials and Methods \label{sec:Materials-and-Methods}}

In SubSec.~(\ref{sub:The-Network-Model}) we describe the spin-glass-like
neural network model whose dynamics we would like to investigate.
Moreover, in SubSec.~(\ref{sub:Plotting-the-Bifurcation-Diagrams})
we introduce a technique for plotting the bifurcation diagram of the
model, provided a set of candidate stationary states and a set of
candidate oscillatory solutions are known. Then, in SubSec.~(\ref{sub:Non-Efficient-Algorithms-for-Generating-the-Stationary-and-Oscillatory-Solutions})
we propose non-optimized brute-force algorithms for generating the
candidate sets. While these algorithms can be applied to networks
with any topology of the synaptic connections, their computational
time increases exponentially with the network size. For this reason,
in SubSec.~(\ref{sub:An-Efficient-Algorithm-for-Generating-the-Stationary-and-Oscillatory-Solutions-in-Sparse-Networks})
we introduce an efficient algorithm that takes advantage of the sparseness
of biological networks for increasing its computational speed. To
conclude, in SubSec.~(\ref{sub:Examples-of-Network-Topologies})
we introduce two standard examples of network topologies, whose bifurcation
structure will be derived in Sec.~(\ref{sec:Results}) through the
use of our algorithms.

\subsection{The Network Model \label{sub:The-Network-Model}}

We study the bifurcation structure of the following spin-glass-like
network model \cite{Fasoli2017}:

\begin{spacing}{0.8}
\begin{center}
{\small{}
\begin{equation}
V_{i}\left(t+1\right)=\frac{1}{M_{i}}\sum_{j=0}^{N-1}J_{ij}\mathscr{H}\left(V_{j}\left(t\right)-\theta_{j}\right)+I_{i},\quad i=0,...,N-1.\label{eq:network-equations}
\end{equation}
}
\par\end{center}{\small \par}
\end{spacing}

\noindent In Eq.~(\ref{eq:network-equations}), $N$ is the number
of neurons in the network, while $V_{i}$ and $I_{i}$ are the membrane
potential and the external stimulus of the $i$th neuron, respectively.
$J_{ij}$ is the synaptic weight from the $j$th (presynaptic) to
the $i$th (postsynaptic) neuron. The collection of the synaptic weights,
for $i,j=0,\ldots,N-1$, defines the synaptic connectivity matrix
$J$, which in this article is supposed to be arbitrary (generally
asymmetric). In Eq.~(\ref{eq:network-equations}), $\mathscr{H}\left(\cdot\right)$
represents the Heaviside step function:

\begin{spacing}{0.8}
\begin{center}
{\small{}
\begin{equation}
\mathscr{H}\left(V-\theta\right)=\begin{cases}
0 & \mathrm{if}\;V\leq\theta\\
\\
1 & \mathrm{otherwise},
\end{cases}\label{eq:Heaviside-step-function}
\end{equation}
}
\par\end{center}{\small \par}
\end{spacing}

\noindent where $\theta$ is the firing threshold. The firing rate
of the $i$th neuron is the binary variable defined as $\nu_{i}\overset{\mathrm{def}}{=}\mathscr{H}\left(V_{i}-\theta_{i}\right)\in\left\{ 0,1\right\} $,
so that $\nu_{i}=0$ if the neuron is not firing, and $\nu_{i}=1$
if it is firing at the maximum rate. Moreover, in Eq.~(\ref{eq:network-equations})
the parameter $M_{i}$ represents the number of presynaptic neurons
that are directly connected to the $i$th (postsynaptic) neuron. Therefore
$M_{i}$ is a normalization factor, that prevents the divergence of
the sum $\sum_{j=0}^{N-1}J_{ij}\mathscr{H}\left(V_{j}\left(t\right)-\theta_{j}\right)$
for $N\gg1$.

The \textit{state-to-state transition probability matrix} $\mathcal{P}$
provides a convenient way to describe the dynamics of the firing rates
$\nu_{i}$. By defining $\boldsymbol{\nu}\left(t\right)\overset{\mathrm{def}}{=}\nu_{0}\left(t\right)\nu_{1}\left(t\right)\ldots\nu_{N-1}\left(t\right)$
as the binary string obtained by concatenating the firing rates at
time $t$, in \cite{Fasoli2017} we proved that the entries of the
$2^{N}\times2^{N}$ matrix $\mathcal{P}$ are:

\begin{spacing}{0.8}
\begin{center}
{\small{}
\begin{equation}
P\left(\boldsymbol{\nu}\left(t+1\right)|\boldsymbol{\nu}\left(t\right)\right)=\frac{1}{2^{N}}\prod_{j=0}^{N-1}\left[1+\left(-1\right)^{\nu_{j}\left(t+1\right)}\mathrm{sgn}\left(\theta_{j}-\frac{1}{M_{j}}\sum_{k=0}^{N-1}J_{jk}\nu_{k}\left(t\right)-I_{j}\right)\right],\label{eq:conditional-probability}
\end{equation}
}
\par\end{center}{\small \par}
\end{spacing}

\noindent where $\mathrm{sgn}\left(\cdot\right)$ is the sign function,
defined as follows:

\begin{spacing}{0.80000000000000004}
\begin{center}
{\small{}
\[
\mathrm{sgn}\left(x\right)=\begin{cases}
-1 & \mathrm{if}\;x<0\\
\\
\hphantom{-}1 & \mathrm{otherwise}.
\end{cases}
\]
}
\par\end{center}{\small \par}
\end{spacing}

\noindent $P\left(\boldsymbol{\nu}\left(t+1\right)|\boldsymbol{\nu}\left(t\right)\right)\in\left\{ 0,1\right\} $
represents the probability of the transition $\boldsymbol{\nu}\left(t\right)\rightarrow\boldsymbol{\nu}\left(t+1\right)$
to occur, for specific firing-rate states $\boldsymbol{\nu}\left(t\right)$
and $\boldsymbol{\nu}\left(t+1\right)$, given the network is in the
state $\boldsymbol{\nu}\left(t\right)$ at the time instant $t$.
For example, if at the time instant $t$ the network is in the state
$\boldsymbol{\nu}\left(t\right)=00\cdots0$ and we want to check if
the neurons flip their firing rates at the next time step (so that
$\boldsymbol{\nu}\left(t+1\right)=11\cdots1$), then $P\left(\boldsymbol{\nu}\left(t+1\right)|\boldsymbol{\nu}\left(t\right)\right)=1$
if the transition $00\cdots0\rightarrow11\cdots1$ occurs from $t$
to $t+1$, and $P\left(\boldsymbol{\nu}\left(t+1\right)|\boldsymbol{\nu}\left(t\right)\right)=0$
otherwise. In SubSec.~(\ref{sub:Plotting-the-Bifurcation-Diagrams})
we will show how to use Eq.~(\ref{eq:conditional-probability}) for
plotting the bifurcation diagram of the network.

\subsection{Plotting the Bifurcation Diagrams \label{sub:Plotting-the-Bifurcation-Diagrams}}

In this subsection we introduce a general way for plotting the bifurcation
diagrams of networks with arbitrary topology of the synaptic connections.
For a given network, its bifurcation diagram is composed of two panels,
namely the \textit{multistability} and the \textit{oscillation diagrams}.
These diagrams provide a complete picture of the relation between
the stationary/oscillatory solutions of the network and the set of
stimuli (see for example the left and right panels of Figs.~(\ref{fig:example-1})
and (\ref{fig:example-2})). In SubSecs.~(\ref{sub:Mulistability-Diagram})
and (\ref{sub:Oscillation-Diagram}) we describe algorithms for calculating
the multistability and the oscillation diagrams, which are implemented
in the supplemental Python scripts ``Multistability\_Diagram.py''
and ``Oscillation\_Diagram.py'' respectively.

Finally, in the case of networks with homogeneous populations, we
superimpose to these diagrams the regions of the stimuli space where
the system undergoes spontaneous intra-population symmetry-breaking.
In these regions, the stationary and oscillatory solutions calculated
by our algorithms show non-homogeneous firing rates within one or
more populations, despite the homogeneity of their neurophysiological
parameters. More generally, given a neuron $i$ in population $\alpha$,
our algorithms detect the formation of spontaneous symmetry-breaking
in networks where all the terms $\frac{1}{M_{i}}\sum_{j\in\beta}J_{ij}$
and $I_{i}$ depend only on the populations $\alpha,\:\beta$, since
this is the most general symmetry condition for the network, according
to Eq.~(\ref{eq:network-equations}). Therefore, networks with homogeneous
parameters $M_{i},\:J_{ij},\:I_{i}$ represent only a special case
of this condition.

\subsubsection{Multistability Diagram \label{sub:Mulistability-Diagram}}

The firing-rate state $\boldsymbol{\nu}$ is stationary if it satisfies
the condition $P\left(\boldsymbol{\nu}|\boldsymbol{\nu}\right)=1$.
From Eq.~(\ref{eq:conditional-probability}) we observe that this
condition holds if:

\begin{spacing}{0.80000000000000004}
\begin{center}
{\small{}
\begin{equation}
\frac{1}{2}\left[1+\left(-1\right)^{\nu_{j}}\mathrm{sgn}\left(\theta_{j}-\frac{1}{M_{j}}\sum_{k=0}^{N-1}J_{jk}\nu_{k}-I_{j}\right)\right]=1,\label{eq:stationary-states-condition}
\end{equation}
}
\par\end{center}{\small \par}
\end{spacing}

\noindent for $j=0,\ldots,N-1$. By inverting Eq.~(\ref{eq:stationary-states-condition})
with respect to the stimulus $I_{j}$, we get that the equation is
satisfied whenever:

\begin{spacing}{0.80000000000000004}
\begin{center}
{\small{}
\begin{equation}
\begin{cases}
I_{j}\leq\mathcal{I}_{j} & \mathrm{if}\;\nu_{j}=0\\
\\
I_{j}>\mathcal{I}_{j} & \mathrm{if}\;\nu_{j}=1.
\end{cases},\quad\mathcal{I}_{j}\overset{\mathrm{def}}{=}\theta_{j}-\frac{1}{M_{j}}\sum_{k=0}^{N-1}J_{jk}\nu_{k}.\label{eq:stationary-states-stimulus-range-0}
\end{equation}
}
\par\end{center}{\small \par}
\end{spacing}

\noindent Generally, if some neurons share the same stimulus, so that
for example the neurons with indexes in the set $\Gamma_{I}$ receive
the same external current $I$, then from Eq.~(\ref{eq:stationary-states-stimulus-range-0})
we get:

\begin{spacing}{0.80000000000000004}
\begin{center}
{\small{}
\begin{equation}
I\in\left(\Lambda_{I},\Xi_{I}\right],\quad\Lambda_{I}\overset{\mathrm{def}}{=}\underset{j\in\Gamma_{I}:\;\nu_{j}=1}{\max}\mathcal{I}_{j},\quad\Xi_{I}\overset{\mathrm{def}}{=}\underset{j\in\Gamma_{I}:\;\nu_{j}=0}{\min}\mathcal{I}_{j}.\label{eq:stationary-states-stimulus-range-1}
\end{equation}
}
\par\end{center}{\small \par}
\end{spacing}

\noindent Eq.~(\ref{eq:stationary-states-stimulus-range-1}) provides
the ranges of the stimuli (if any) where the firing-rate state $\boldsymbol{\nu}$
is stationary. By calculating the ranges $\left(\Lambda_{I},\Xi_{I}\right]$
for every set $\Gamma_{I}$, we get a complete picture of the relation
between the stationary states and the set of stimuli (see for example
the left panels of Figs.~(\ref{fig:example-1}) and (\ref{fig:example-2})).
If the ranges corresponding to $\mathcal{M}$ different states $\boldsymbol{\nu}$
overlap, the overlapping area has multistability degree $\mathcal{M}$.
In other words, for combinations of stimuli lying in that area, the
network has $\mathcal{M}$ stationary firing rates. Therefore Eq.~(\ref{eq:stationary-states-stimulus-range-1})
allows the analytical derivation of the multistability diagram, extending
the analysis we performed in \cite{Fasoli2017} to networks with arbitrary
topology and size.

Note that Eq.~(\ref{eq:stationary-states-stimulus-range-1}) implies
the calculation of the set $\left(\Lambda_{I},\Xi_{I}\right]$ for
each (potentially) stationary firing-rate state $\boldsymbol{\nu}$.
By defining $\mathscr{S}$ as the set of the candidate stationary
states to be checked, we observe that $\boldsymbol{\nu}\in\mathscr{S}$
is actually stationary if $\Lambda_{I}<\Xi_{I}$. In SubSecs.~(\ref{sub:Non-Efficient-Algorithms-for-Generating-the-Stationary-and-Oscillatory-Solutions})
and (\ref{sub:An-Efficient-Algorithm-for-Generating-the-Stationary-and-Oscillatory-Solutions-in-Sparse-Networks})
we will introduce two different algorithms for generating $\mathscr{S}$,
which are called by the Python script ``Multistability\_Diagram.py''.
Depending on the algorithm, the cardinality $\left|\mathscr{S}\right|$
may be different for a given network topology. This will affect the
overall speed with which the multistability diagram is plotted.

\subsubsection{Oscillation Diagram \label{sub:Oscillation-Diagram}}

In principle, the same approach discussed in the previous subsection
for the multistability diagram of the network can be extended to the
study of neural oscillations. A sequence $\mathcal{O}$, defined as
$\boldsymbol{\nu}\left(0\right)\rightarrow\boldsymbol{\nu}\left(1\right)\rightarrow\ldots\rightarrow\boldsymbol{\nu}\left(\mathcal{T}\right)$
with $\boldsymbol{\nu}\left(0\right)=\boldsymbol{\nu}\left(\mathcal{T}\right)$,
is an actual oscillatory solution (of period $\mathcal{T}\in\left\{ 2,\ldots,2^{N}\right\} $)
if every transition $\boldsymbol{\nu}\left(t\right)\rightarrow\boldsymbol{\nu}\left(t+1\right)$
in $\mathcal{O}$ satisfies the condition $P\left(\boldsymbol{\nu}\left(t+1\right)|\boldsymbol{\nu}\left(t\right)\right)=1$.
Similarly to SubSec.~(\ref{sub:Mulistability-Diagram}), if some
neurons share the same stimulus, so that the neurons with indexes
in the set $\Gamma_{I}$ receive the same external current $I$, then
the sequence $\mathcal{O}$ is an oscillatory solution of the network
dynamics if:

\begin{spacing}{0.80000000000000004}
\begin{center}
{\small{}
\begin{equation}
I\in\left(\Phi_{I},\Psi_{I}\right],\quad\Phi_{I}\overset{\mathrm{def}}{=}\underset{t\in\mathscr{T}}{\max}\left(\underset{j\in\Gamma_{I}:\;\nu_{j}\left(t+1\right)=1}{\max}\mathcal{I}_{j}\left(t\right)\right),\quad\Psi_{I}\overset{\mathrm{def}}{=}\underset{t\in\mathscr{T}}{\min}\left(\underset{j\in\Gamma_{I}:\;\nu_{j}\left(t+1\right)=0}{\min}\mathcal{I}_{j}\left(t\right)\right),\label{eq:oscillations-stimulus-range}
\end{equation}
}
\par\end{center}{\small \par}
\end{spacing}

\noindent where $\mathscr{T}\overset{\mathrm{def}}{=}\left\{ 0,\ldots,\mathcal{T}-1\right\} $
and $\mathcal{I}_{j}\left(t\right)\overset{\mathrm{def}}{=}\theta_{j}-\frac{1}{M_{j}}\sum_{k=0}^{N-1}J_{jk}\nu_{k}\left(t\right)$.
Finally, by calculating the ranges $\left(\Phi_{I},\Psi_{I}\right]$
for every set $\Gamma_{I}$, we get a complete analytical picture
of the relation between the oscillatory solutions and the set of stimuli
(see for example the right panels of Figs.~(\ref{fig:example-1})
and (\ref{fig:example-2})).

Similarly to the case of the multistability diagram, we observe that
Eq.~(\ref{eq:oscillations-stimulus-range}) implies the calculation
of the set $\left(\Phi_{I},\Psi_{I}\right]$ for each (potential)
oscillatory solution. In SubSecs.~(\ref{sub:Non-Efficient-Algorithms-for-Generating-the-Stationary-and-Oscillatory-Solutions})
and (\ref{sub:An-Efficient-Algorithm-for-Generating-the-Stationary-and-Oscillatory-Solutions-in-Sparse-Networks})
we will introduce two different algorithms for generating the set
$\mathscr{O}$ of the candidate oscillatory solutions, which are called
by the Python script ``Oscillation\_Diagram.py''.

\subsection{Non-Efficient Algorithms for Generating the Stationary and Oscillatory
Solutions \label{sub:Non-Efficient-Algorithms-for-Generating-the-Stationary-and-Oscillatory-Solutions}}

In SubSec.~(\ref{sub:Plotting-the-Bifurcation-Diagrams}) we described
an algorithm for plotting the multistability and oscillation diagrams,
provided some sets of candidate stationary and oscillatory solutions,
$\mathscr{S}$ and $\mathscr{O}$ respectively, are known. In the
present section we introduce non-optimized brute-force algorithms,
implemented in the Python script ``Non\_Efficient\_Algorithm.py'',
that generate the sets $\mathscr{S}$ and $\mathscr{O}$ for networks
with arbitrary topology.

\subsubsection{Generation of the Set $\mathscr{S}$ \label{sub:Generation-of-the-Set-S-Arbitrary-Networks}}

The simplest way for generating the set $\mathscr{S}$ is to fill
it with all the $2^{N}$ states of the firing rates, from $00\cdots0$
to $11\cdots1$. This approach allows the evaluation of the multistability
diagram of small-size networks, since its computational time increases
exponentially with $N$. We observe that only the states in $\mathscr{S}$
that satisfy the constraint $\Lambda_{I}<\Xi_{I}$ (see SubSec.~(\ref{sub:Plotting-the-Bifurcation-Diagrams}))
are actually stationary, therefore the set $\mathscr{S}$ generated
by this algorithm is usually oversized. In SubSec.~(\ref{sub:Future-Directions})
we will discuss a way to reduce $\left|\mathscr{S}\right|$ for specific
network topologies.

\subsubsection{Generation of the Set $\mathscr{O}$ \label{sub:Generation-of-the-Set-O-Arbitrary-Networks}}

Unfortunately, the brute-force approach described above for the generation
of the set $\mathscr{S}$ is unfeasible for the study of neural oscillations,
due to computational time. Indeed, the number of possible oscillations
in a network of size $N$ is $n_{\mathcal{O}}=\sum_{k=2}^{2^{N}}\binom{2^{N}}{k}\left(k-1\right)!$,
which grows as $\sim\left(2^{N}-1\right)!$ for $N\rightarrow\infty$
(see Appx.~(\hyperref[sec:Appendix-A]{A})). The best solution we
found to reduce the computational time, given a network with arbitrary
topology, is to obtain $\mathscr{O}$ by discretizing the stimuli
space, and then by solving iteratively Eq.~(\ref{eq:network-equations})
for every combination of stimuli and for every initial condition of
the firing rates. In other words, we discretized the stimuli space
through a grid $G$ composed of $n_{G}$ points, each one representing
a combination of stimuli. Then we solved iteratively the equation
$V_{i}\left(t+1\right)=\frac{1}{M_{i}}\sum_{j=0}^{N-1}J_{ij}\nu_{j}\left(t\right)+I_{i}$,
with $\nu_{j}\left(t\right)=\mathscr{H}\left(V_{j}\left(t\right)-\theta_{i}\right)$,
for each of the $n_{G}$ combinations of stimuli of the grid $G$,
and for each of the $2^{N}$ initial conditions $\boldsymbol{\nu}\left(0\right)$.
If the firing-rate vector $\boldsymbol{\nu}\left(t\right)$ calculated
through Eq.~(\ref{eq:network-equations}) oscillates according to
a sequence $\mathcal{O}$ for a given initial condition $\boldsymbol{\nu}\left(0\right)$,
then $\mathcal{O}\in\mathscr{O}$. Finally, we filtered the set $\mathcal{O}$
in order to remove duplicate oscillations, so that the set will contain
every oscillation exactly once (for example, in a $2$-neurons circuit,
the oscillations $01\rightarrow10\rightarrow11\rightarrow01$ and
$10\rightarrow11\rightarrow01\rightarrow10$ are circularly identical,
therefore one must be discarded). Through this approach, we can derive
the actual oscillations in the set $\mathscr{O}$, by solving Eq.~(\ref{eq:network-equations})
$2^{N}n_{G}$ times. Generally $2^{N}n_{G},\,\left|\mathscr{O}\right|\ll n_{\mathcal{O}}$,
so that if we calculate the range $\left(\Phi_{I},\Psi_{I}\right]$
through Eq.~(\ref{eq:oscillations-stimulus-range}) for every oscillation
in $\mathscr{O}$, the computational time required for deriving the
oscillation diagram now increases as $2^{N}$ with the network size,
rather than $\left(2^{N}-1\right)!$.

However, while being much faster, this algorithm does not guarantee
that the resulting oscillation diagram is complete. Indeed, for some
oscillations, the range $\left(\Phi_{I},\Psi_{I}\right]$ could be
smaller than the grid resolution. For this reason, the parameter $n_{G}$
must be chosen accurately in order to avoid relevant information loss
(which may occur when $n_{G}$ is too small for a given network) and
an excessive computational load of the algorithm (which may occur
when $n_{G}$ is too large for the computational power available).

To conclude, we observe that the oscillation diagram may be calculated
in a purely numerical way, by solving Eq.~(\ref{eq:network-equations})
for every combination of stimuli on the grid, without making use of
the analytical formula (\ref{eq:oscillations-stimulus-range}). However,
by relying on Eq.~(\ref{eq:oscillations-stimulus-range}), our semi-analytical
approach allows the analytical derivation of the oscillation diagram,
which would not be possible otherwise. Moreover, it is easy to verify
that the use of Eq.~(\ref{eq:oscillations-stimulus-range}) allows
a reduction of the resolution of the grid $G$ compared to a purely
numerical approach, resulting in a much faster derivation of the oscillation
diagram.

\subsection{An Efficient Algorithm for Generating the Stationary and Oscillatory
Solutions in Sparse Networks \label{sub:An-Efficient-Algorithm-for-Generating-the-Stationary-and-Oscillatory-Solutions-in-Sparse-Networks}}

We observe that the algorithms described in SubSec.~(\ref{sub:Non-Efficient-Algorithms-for-Generating-the-Stationary-and-Oscillatory-Solutions})
can be applied to networks with any topology of the synaptic connections
(dense or sparse), but their computational time increases as $2^{N}$
with the network size. In this subsection we show that, for the specific
case of sparse networks, it is possible to take advantage of the sparseness
of the synaptic connections to define an efficient algorithm for the
generation of the sets $\mathscr{S}$ and $\mathscr{O}$. This approach,
which we implemented in the Python script ``Sparse\_Efficient\_Algorithm.py'',
may outperform of several orders of magnitude the non-optimized approaches
introduced in SubSec.~(\ref{sub:Non-Efficient-Algorithms-for-Generating-the-Stationary-and-Oscillatory-Solutions}).

\subsubsection{Generation of the Set $\mathscr{S}$ \label{sub:Generation-of-the-Set-S-Sparse-Networks}}

As we saw in SubSec.~(\ref{sub:Mulistability-Diagram}), in order
to identify the stationary states of the network we need to check
if the condition (\ref{eq:stationary-states-condition}) is satisfied
for each neuronal index $j=0,\ldots,N-1$, given a set of stimuli
$I_{0},\ldots,I_{N-1}$. If the $j$th neuron receives inputs from
all the other neurons in the network (i.e. if $J_{jk}\neq0$ for $k=0,\ldots,N-1$),
the condition (\ref{eq:stationary-states-condition}) must be checked
for all the $2^{N}$ binary states $\boldsymbol{\nu}$ of length $N$,
from $00\cdots0$ to $11\cdots1$. On the contrary, in sparse networks
some synaptic weights are equal to zero. If some neurons do not project
a synaptic connection to the $j$th neuron (i.e. if $J_{jk}=0$ for
some index $k$), the corresponding firing rates $\nu_{k}$ will not
affect the sum $\sum_{k=0}^{N-1}J_{jk}\nu_{k}$. Therefore for sparse
networks there is no need to check Eq.~(\ref{eq:stationary-states-condition})
for all the binary states of length $N$, unlike the case of dense
networks. This observation allows us to introduce an efficient algorithm
for finding the candidate stationary states of sparse neural networks,
which is described below.

At step $0$, we set $j=0$ and we call $P_{0}$ the set of neurons
with indexes $k\neq0$ that do not project a synaptic connection to
the $0$th neuron (i.e. $k\in P_{0}$ if $J_{0k}=0$ and $k\neq0$).
Moreover, we call $Q_{0}$ the set of the remaining $N-\left|P_{0}\right|$
neurons. This is the set of neurons that can affect the condition
(\ref{eq:stationary-states-condition}) through their firing rates.
In particular, the $0$th neuron belongs to $Q_{0}$ since it can
affect the condition (\ref{eq:stationary-states-condition}) through
the term $\left(-1\right)^{\nu_{0}}$ even in the case when $J_{00}=0$
(therefore $\left|Q_{0}\right|=M_{0}$ if $J_{00}\neq0$, and $\left|Q_{0}\right|=M_{0}+1$
if $J_{00}=0$, where $M_{0}$ is the incoming vertex degree of the
$0$th neuron, see Eq.~(\ref{eq:network-equations})). At this step,
we only need to check Eq.~(\ref{eq:stationary-states-condition})
for all the binary states $s_{0}$ of length $\left|Q_{0}\right|$.
If no state $s_{0}$ satisfies Eq.~(\ref{eq:stationary-states-condition}),
the network has no stationary solution and the algorithm is stopped.
Otherwise we call $\mathscr{S}$ the set of states $s_{0}$ that satisfy
Eq.~(\ref{eq:stationary-states-condition}), and we switch to the
neuronal index $j=1$, as described below.

At step $1$, we set $j=1$ and we call $P_{1}$ the set of neurons
with indexes $k\neq1$ that do not project to the $1$st neuron. Moreover
we call $Q_{1}$ the set of the remaining $N-\left|P_{1}\right|$
neurons, and $R_{1}=Q_{1}\cap Q_{0}$. Then we generate all the binary
states $s_{1}$ of length $\left|Q_{1}\right|-\left|R_{1}\right|$,
and we use each of them to complete the states in $\mathscr{S}$,
by creating new binary states of length $\left|Q_{0}\right|+\left|Q_{1}\right|-\left|R_{1}\right|$.
For example, we suppose that at step $0$ we got $Q_{0}=\left\{ 0,\:7,\:8\right\} $
and $\mathscr{S}=\left\{ 010,\:111\right\} $, while at step $1$
we got $Q_{1}=\left\{ 1,\:2\right\} $. Then the states $s_{1}=00,$
$01$, $10$, $11$ will be used to fill the states in $\mathscr{S}$
according to the index $k$, generating the new states $0\boldsymbol{0}\boldsymbol{0}10$,
$0\boldsymbol{0}\boldsymbol{1}10$, $0\boldsymbol{1}\boldsymbol{0}10$,
$0\boldsymbol{1}\boldsymbol{1}10$ and $1\boldsymbol{0}\boldsymbol{0}11$,
$1\boldsymbol{0}\boldsymbol{1}11$, $1\boldsymbol{1}\boldsymbol{0}11$,
$1\boldsymbol{1}\boldsymbol{1}11$ (the bits of the states $s_{1}$
are highlighted in bold). Then the algorithm checks the condition
(\ref{eq:stationary-states-condition}) for $j=1$ on the newly generated
states (when $Q_{1}$ is empty, the algorithm checks the condition
directly on the states in $\mathscr{S}$). If no state satisfies Eq.~(\ref{eq:stationary-states-condition}),
then the network has no stationary solution, therefore the set $\mathscr{S}$
is cleared and the algorithm is stopped. Otherwise $\mathscr{S}$
is cleared and filled with the newly generated states that satisfy
Eq.~(\ref{eq:stationary-states-condition}). Then we proceed to the
neuronal index $j=2$.

In a similar way, for $2\leq j\leq N-1$, we define $P_{j}$ as the
set of neurons with indexes $k\neq j$ which do not project to the
$j$th neuron. Moreover we call $Q_{j}$ the set of the remaining
$N-\left|P_{j}\right|$ neurons, and $R_{j}=Q_{j}\cap\left(\bigcup_{n=0}^{j-1}Q_{n}\right)$.
The procedure described at step $1$ is repeated iteratively through
the steps $2$, $3$, ... by completing the states in $\mathscr{S}$
through the binary strings $s_{j}$ of length $\left|Q_{j}\right|-\left|R_{j}\right|$.
If the algorithm does not stop, when the $j$th step has been completed
the states in $\mathscr{S}$ have length $L_{j}=\sum_{n=0}^{j}\left|Q_{n}\right|-\sum_{n=1}^{j}\left|R_{n}\right|$.
We observe that $L_{j}$ quickly tends to $N$ for increasing $j$.
For example, if the network has a ring topology (i.e. if each neuron
receives a connection only from the previous neuron and projects a
connection only to the next one), we get $\left|Q_{n}\right|=2$ for
$n=0,\ldots,N-1$, $\left|R_{n}\right|=1$ for $n=1,\ldots,N-2$ and
$\left|R_{N-1}\right|=2$, so that $L_{j}\rightarrow N$ for $j\rightarrow N-1$.

At the end of the process, $\mathscr{S}$ will contain the actual
stationary states of the network (if any), namely all the states $\boldsymbol{\nu}$
of length $N$ that satisfy the condition $P\left(\boldsymbol{\nu}|\boldsymbol{\nu}\right)=1$,
if the stimuli $I_{0},\ldots,I_{N-1}$ are all known. The efficiency
of the algorithm is directly proportional to the sparseness of the
matrix $J$ and inversely proportional to the number of stationary
states of the network. Moreover, further speed-up is achieved by sorting
the neuronal indexes with ascending vertex degrees $M_{j}$, before
running the algorithm. This is due to the fact that the algorithm
slows down when both $\left|Q_{j}\right|-\left|R_{j}\right|$ and
$\left|\mathscr{S}\right|$ are large. However, we observe that if
at some step $\overline{j}$ a string of length $L_{\overline{j}}$
does not satisfy the condition (\ref{eq:stationary-states-condition}),
then during the next steps the algorithm will not check all the $2^{N-L_{\overline{j}}}$
binary states of length $N$ that contain that string. Since $L_{\overline{j}}\approx\sum_{n=0}^{\overline{j}}M_{n}-\sum_{n=1}^{\overline{j}}\left|R_{n}\right|$,
if the vertex degrees $M_{n}$ are small $\forall n\leq\overline{j}$,
then the string length $L_{\overline{j}}$ must be small. For this
reason, the number of (non-stationary) states containing the string
of length $L_{\overline{j}}$ is large ($2^{N-L_{\overline{j}}}$).
Being non-stationary, this large number of states will not fill the
set $\mathscr{S}$, which therefore remains small. When the algorithm
will proceed by iterating over the indexes $j$ with large degree
$M_{j}$ (which are computationally more expensive, since the algorithm
has to iterate over all the binary states of length $\left|Q_{j}\right|-\left|R_{j}\right|\approx M_{j}-\left|R_{j}\right|$),
the number of states in $\mathscr{S}$ will be small. This reduces
the computational load in the slowest part of the algorithm, resulting
in an overall speed-up. On the contrary, for the same reason the algorithm
slows down when the neurons are sorted with descending $M_{j}$.

\begin{figure}
\begin{centering}
\includegraphics[scale=0.36]{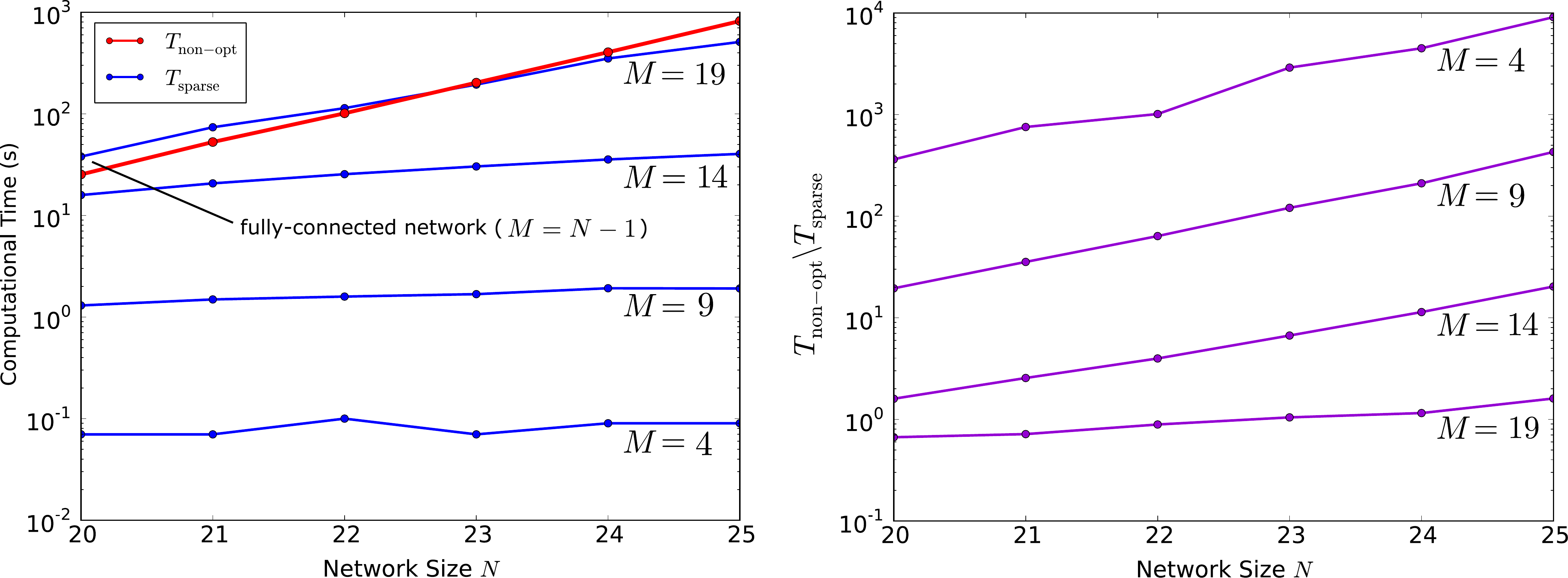}
\par\end{centering}

\protect\caption{\label{fig:speed-test} \small\textbf{ Speed test of the algorithm
for sparse networks.} This figure shows the improvement of performance
achieved by the algorithm for sparse networks described in SubSec.~(\ref{sub:Generation-of-the-Set-S-Sparse-Networks}),
compared to the non-optimized algorithm of SubSec.~(\ref{sub:Generation-of-the-Set-S-Arbitrary-Networks}).
We tested the algorithms by calculating the stationary states of a
sparse circulant network with topology $J=\overline{J}\mathrm{circ}\left(0,1,\ldots,1,0,\ldots,0\right)$.
Each neuron has incoming vertex degree $M\in\left[0,N-1\right]$,
so that the sparseness of the network is proportional to $M$, according
to the formula $\frac{\#\,\mathrm{synapses}}{N^{2}-N}=\frac{M}{N-1}\in\left[0,1\right]$.
Moreover, we set the overall synaptic strength to $\overline{J}=10$
and the external stimuli to $I_{0}=\ldots=I_{N-1}=0$. For these values
of the parameters, the network has two stationary states, $\forall M,\:N$
considered in this figure. The left panel shows the computational
times $T_{\mathrm{sparse}}$ and $T_{\mathrm{non-opt}}$ (see text)
that are required for calculating the stationary states by means of
an Intel\textsuperscript{\textregistered} Core\texttrademark{} i5-5300U
CPU clocked at 2.30GHz with 16 GB RAM. In particular, we observe that
$T_{\mathrm{sparse}}\ll T_{\mathrm{non-opt}}$ if $M$ is sufficiently
smaller that $N-1$, while for small, highly dense networks the non-optimized
algorithm may outperform the algorithm for sparse networks. Moreover,
note that $T_{\mathrm{non-opt}}$ does not depend on $M$. The right
panel shows the ratio $T_{\mathrm{non-opt}}\backslash T_{\mathrm{sparse}}$,
namely the speed gain of the algorithm for sparse networks with respect
to the non-optimized one, as a function of $N$ and $M$. Similar
results hold for the calculation of the oscillatory solutions, when
$T_{\mathrm{sparse}}$ is evaluated through the algorithm introduced
in SubSec.~(\ref{sub:Generation-of-the-Set-O-Sparse-Networks}) for
$\mathcal{T}_{\mathrm{max}}\ll\frac{N}{\max_{j}\left(M_{j}-\left|R_{j}\right|\right)}$
(see text). For increasing values of $\mathcal{T}_{\mathrm{max}}$
the algorithm for sparse networks becomes less and less efficient
compared to the non-optimized algorithm of SubSec.~(\ref{sub:Generation-of-the-Set-O-Arbitrary-Networks})
(results not shown).}

\end{figure}

Fig.~(\ref{fig:speed-test}) shows a comparison between the computational
time $T_{\mathrm{sparse}}$ required by our algorithm for calculating
the stationary states of a given network topology for a fixed set
of stimuli $I_{0},\ldots,I_{N-1}$, and the computational time $T_{\mathrm{non-opt}}$
required for performing the same calculation through the non-optimized
brute-force approach introduced in SubSec.~(\ref{sub:Generation-of-the-Set-S-Arbitrary-Networks}).
 The figure shows that the speed gain $\frac{T_{\mathrm{non-opt}}}{T_{\mathrm{sparse}}}$
of the algorithm for sparse networks with respect to the non-optimized
algorithm is generally much larger than $1$, and it increases with
the network size and sparseness. The computational time of the non-optimized
algorithm increases exponentially as $2^{N}$ regardless of the network
sparseness. However, for small and highly dense networks this algorithm
could outperform that for sparse networks ($\frac{T_{\mathrm{non-opt}}}{T_{\mathrm{sparse}}}<1$).

\begin{figure}
\begin{centering}
\includegraphics[scale=0.25]{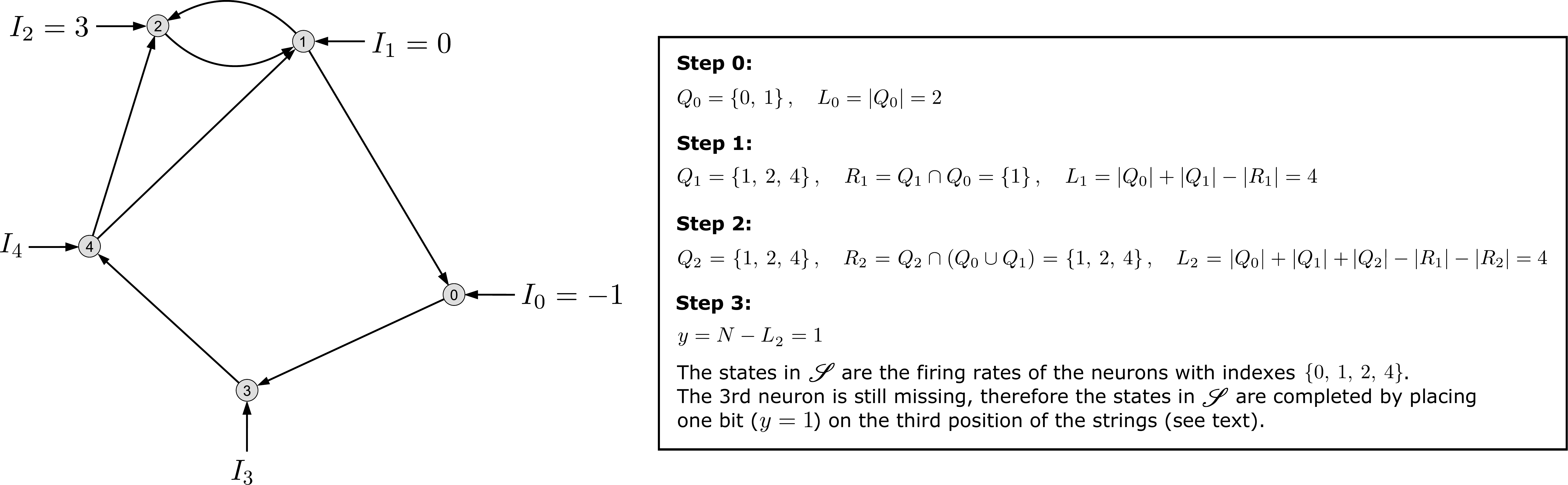}
\par\end{centering}

\protect\caption{\label{fig:algorithm-progression} \small\textbf{ An example of the
progression of the efficient algorithm for a specific connectivity
matrix.} The left panel shows the directed graph of the connectivity
matrix, so that each node represents a neuron in the network, while
a black arrow from a node $j$ to a node $i$ represents a synaptic
connection with weight $J_{ij}$. Moreover, each node $i$ receives
an external input current $I_{i}$. In the specific case considered
in this figure, the inputs to the neurons $0$, $1$, $2$ are known
(i.e. $I_{0}=-1$, $I_{1}=0$, $I_{2}=3$), and we want to derive
the multistability diagram with respect to the unspecified inputs
$I_{3}$ and $I_{4}$. The right panel shows the progression of the
efficient algorithm for sparse networks during the evaluation of the
set of the candidate stationary states, $\mathscr{S}$. In the first
three steps, the algorithm determines the states that satisfy the
condition (\ref{eq:stationary-states-condition}) for $j=0,\,1,\,2$,
since the currents $I_{0,1,2}$ are known. In other words, the algorithm
determines the binary strings, composed of the firing rates of the
neurons $0$, $1$, $2$ (since their external currents are known)
and of the firing rates of the neurons that project a synaptic connection
to the nodes $0$, $1$, $2$ (i.e. the neuron $4$ in this example),
to be put in $\mathscr{S}$. These strings of length $4$ do not represent
yet the candidate stationary states of the network, since the firing
rate of the $3$rd neuron is still missing. Therefore in the final
step the strings in $\mathscr{S}$ are completed by inserting the
bit of the $3$rd neuron. We observe that, unlike for the neurons
$0$, $1$, $2$, the condition (\ref{eq:stationary-states-condition})
cannot be checked for the $3$rd neuron, since the current $I_{3}$
is unspecified. For this reason, each string in $\mathscr{S}$ will
give rise to two candidate stationary states, one for each possible
value ($0$ or $1$) of the firing rate of the $3$rd neuron (see
text). At the end of this process, $\mathscr{S}$ will contain candidate
stationary states $\boldsymbol{\nu}$ of length $5$, which will be
used by the script ``Multistability\_Diagram.py''. Finally, the
latter will derive the multistability diagram by inverting the relation
$P\left(\boldsymbol{\nu}|\boldsymbol{\nu}\right)=1$ in the unspecified
currents $I_{3}$ and $I_{4}$, for all the states $\boldsymbol{\nu}$
in $\mathscr{S}$. This technique will be extended to the study of
neural oscillations, in SubSec.~(\ref{sub:Generation-of-the-Set-O-Sparse-Networks}).}
\end{figure}

As we said, when the currents $I_{0},\ldots,I_{N-1}$ are all known,
the algorithm for sparse networks generates a set $\mathscr{S}$ containing
the actual stationary states of the network. However, since our purpose
is to plot the multistability diagram, we observe that the values
of the currents that define the parameter space of the diagram are
generally not fixed (see e.g. Fig.~(\ref{fig:algorithm-progression}),
where the currents $I_{3}$ and $I_{4}$ are unspecified since they
represent the bifurcation parameters of the network). For this reason,
in order to derive the multistability diagram, the algorithm described
above must be adapted for generating a set $\mathscr{S}$ whenever
some inputs are unspecified. In what follows we propose two different
solutions.

The first is to apply our algorithm for all the combinations of the
bifurcation stimuli on a grid in the stimuli space, similarly to the
method described in SubSec.~(\ref{sub:Generation-of-the-Set-O-Arbitrary-Networks})
for the study of oscillations. This approach works for any set of
stimuli (e.g. when all the neurons share the same stimulus or when
each neuron receives a distinct external input), but it does not guarantee
that the resulting multistability diagram is complete if the resolution
of the grid is not high enough.

Unlike the first method described above, the second way to use our
algorithm (which is the one we implemented in the Python script ``Sparse\_Efficient\_Algorithm.py'')
is specific to systems where the external currents that we vary during
the bifurcation analysis are injected into a limited number of neurons.
In other words, we suppose that the stimuli that span the bifurcation
diagram are $I_{x},\ldots,I_{N-1}$, where $x$ is an integer close
or equal to $N-1$ (see e.g. the network considered in Eq.~(\ref{eq:network-structure-example-2}),
where $x=N-2$ after a proper rearrangement of the neural indexes).
We also suppose that the remaining stimuli, i.e. $I_{0},\ldots,I_{x-1}$,
are known and fixed, so that the algorithm for sparse networks can
be applied for $j=0,\ldots,x-1$. The algorithm will generate a set
$\mathscr{S}$ containing incomplete stationary states, namely binary
strings of length $L_{x-1}$. Now, the $2^{y}$ binary states of length
$y\overset{\mathrm{def}}{=}N-L_{x-1}$ from $00\cdots0$ to $11\cdots1$
must be used to complete the states in $\mathscr{S}$, creating $K=2^{y}\left|\mathscr{S}\right|$
binary strings of length $N$. For example, in the case of the $5$-neurons
network shown in Fig.~(\ref{fig:algorithm-progression}), we suppose
we got $\mathscr{S}=\left\{ 1001,\:1101\right\} $ after the step
$x-1$ (with $x=3$) has been completed, and that the states in $\mathscr{S}$
are the firing rates of the neurons with indexes $\left\{ 0,\:1,\:2,\:4\right\} $.
Then $y=1$, therefore the complete binary strings are $\boldsymbol{\nu}=100\boldsymbol{0}1,\:100\boldsymbol{1}1,\:110\boldsymbol{0}1,\:110\boldsymbol{1}1$
(the bit corresponding to $y$ is highlighted in bold). 
 These are candidate stationary states, which can be used to derive
the multistability diagram by inverting the relation $P\left(\boldsymbol{\nu}|\boldsymbol{\nu}\right)=1$
in the currents $I_{x},\ldots,I_{N-1}$. Note that the relation between
$x$ and $y$ depends on the topology of the network. However, if
$x$ is close or equal to $N-1$, then $y$ is generally small, because
most of the binary digits in every stationary firing-rate state are
determined, through our algorithm, by the known currents $I_{0},\ldots,I_{x-1}$.
In this case, the overall efficiency of the algorithm will be high
for large $N$, since $K$ does not diverge exponentially with the
network size.

\subsubsection{Generation of the Set $\mathscr{O}$ \label{sub:Generation-of-the-Set-O-Sparse-Networks}}

The algorithm introduced in SubSec.~(\ref{sub:Generation-of-the-Set-S-Sparse-Networks})
can be naturally extended to find the oscillatory solutions of the
network, as we describe below. In SubSec.~(\ref{sub:Oscillation-Diagram})
we observed that the sequence $\boldsymbol{\nu}\left(0\right)\rightarrow\boldsymbol{\nu}\left(1\right)\rightarrow\ldots\rightarrow\boldsymbol{\nu}\left(\mathcal{T}\right)$,
with $\boldsymbol{\nu}\left(0\right)=\boldsymbol{\nu}\left(\mathcal{T}\right)$,
is an oscillatory solution if every transition $\boldsymbol{\nu}\left(t\right)\rightarrow\boldsymbol{\nu}\left(t+1\right)$
satisfies the condition $P\left(\boldsymbol{\nu}\left(t+1\right)|\boldsymbol{\nu}\left(t\right)\right)=1$.
For this reason, the initially empty set $\mathscr{O}$ of the oscillatory
solutions with fixed period $\mathcal{T}$ is populated as follows.
First, at the $j$th step of the algorithm (for $j=0,\ldots,N-1$),
we need to generate all the binary states of length $\left|Q_{j}\right|-\left|R_{j}\right|$
(where $Q_{j}$ and $R_{j}$ are defined as in SubSec.~(\ref{sub:Generation-of-the-Set-S-Sparse-Networks}),
and $R_{0}\overset{\mathrm{def}}{=}\left\{ \right\} $) for each time
instant $t=0,\ldots,\mathcal{T}-1$. In other words, the algorithm
creates all the $2^{\mathcal{T}\left(\left|Q_{j}\right|-\left|R_{j}\right|\right)}$
$\mathcal{T}$-tuples of states of length $\left|Q_{j}\right|-\left|R_{j}\right|$.
Then, similarly to SubSec.~(\ref{sub:Generation-of-the-Set-S-Sparse-Networks}),
we have to use these states for completing those in the set $\mathscr{O}$,
and to check if the states in $\mathscr{O}$ satisfy the condition
$\frac{1}{2}\left[1+\left(-1\right)^{\nu_{j}\left(t+1\right)}\mathrm{sgn}\left(\theta_{j}-\frac{1}{M_{j}}\sum_{k=0}^{N-1}J_{jk}\nu_{k}\left(t\right)-I_{j}\right)\right]=1$
for $t=0,\ldots,\mathcal{T}-1$. At the end of the $j$th step, the
set $\mathscr{O}$, if not empty, is cleared and filled with the newly
generated states that satisfy the condition, before proceeding to
the $\left(j+1\right)$th step of the algorithm. The algorithm is
stopped only if $\mathscr{O}$ is empty or after the $\left(N-1\right)$th
step is completed. Then, similarly to the method of SubSec.~(\ref{sub:Generation-of-the-Set-O-Arbitrary-Networks}),
we discard the duplicate oscillations in the set $\mathscr{O}$ .
At the end of the process, if the currents $I_{0},\ldots,I_{N-1}$
are all known, the set will contain exactly once all the actual oscillatory
solutions of the given network topology with fixed period $\mathcal{T}$
. Whenever some external stimuli are unspecified, the algorithm is
adapted as we discussed in SubSec.~(\ref{sub:Generation-of-the-Set-S-Sparse-Networks})
for the stationary states.

We observe that this approach, which we implemented in the script
``Sparse\_Efficient\_Algorithm.py'', detects only the oscillations
with a fixed period $\mathcal{T}$. Therefore this procedure must
be repeated for $\mathcal{T}=2,\ldots,\mathcal{T}_{\mathrm{max}}$,
where the maximum period $\mathcal{T}_{\mathrm{max}}$ is a user-defined
parameter, that should be chosen according to the computational power
available. Generally, for $\mathcal{T}_{\mathrm{max}}<2^{N}$ the
algorithm is not guaranteed to find all the oscillatory solutions.
However, setting $\mathcal{T}_{\mathrm{max}}$ close to or larger
than $Z\overset{\mathrm{def}}{=}\frac{N}{\max_{j}\left(M_{j}-\left|R_{j}\right|\right)}$
results in an overall (at least exponential with $N$) slowing down
of the algorithm, due to the generation of the $2^{\mathcal{T}\left(\left|Q_{j}\right|-\left|R_{j}\right|\right)}\approx2^{\mathcal{T}\left(M_{j}-\left|R_{j}\right|\right)}$
$\mathcal{T}$-tuples. Therefore, similarly to SubSec.~(\ref{sub:Generation-of-the-Set-O-Arbitrary-Networks}),
the study of oscillations proves to be computationally expensive,
requiring a compromise between precision and computational speed,
which in this case is generally achieved by setting $2\ll\mathcal{T}_{\mathrm{max}}\ll Z$
for large $Z$. We observe that $Z$ increases with the network sparseness
for a fixed network size $N$, therefore in sparse networks the algorithm
will detect more oscillation periods in a fixed amount of time, compared
to dense networks. Unfortunately, it is not possible to know a priori
the effective maximum period of the oscillations generated by a network.
However, in the sparse topologies we analyzed, this was usually small,
and oscillations with period equal to or larger than the network size
appear much more rarely (see e.g. the right panels of Fig.~(\ref{fig:example-2}),
where random sparse networks of size $N=4,\,6,\,8$ showed only oscillations
with period $2$ and $4$). Typically, oscillation periods larger
than $N$ occur in systems with built-on-purpose, and usually very
regular, topologies.

To conclude, we observe that more generally this technique can be
used to calculate efficiently the whole state-to-state transition
probability matrix $\mathcal{P}$. According to Eq.~(\ref{eq:conditional-probability}),
this is defined as the $2^{N}\times2^{N}$ binary matrix of the conditional
probabilities $P\left(\cdot|\cdot\right)$ of the firing rates. More
precisely, $\mathcal{P}_{ij}=P\left(\boldsymbol{b}_{N}\left(i\right)|\boldsymbol{b}_{N}\left(j\right)\right)\in\left\{ 0,1\right\} $,
for $i,\:j=0,\ldots,2^{N}-1$, where the firing rate $\boldsymbol{b}_{N}\left(k\right)$
is the binary representation of length $N$ of the index $k$ (e.g.,
if $N=5$, then $\boldsymbol{b}_{N}\left(14\right)=01110$). Our algorithm
provides all the firing-rate pairs $\left(\boldsymbol{b}_{N}\left(i\right),\boldsymbol{b}_{N}\left(j\right)\right)$
such that $P\left(\boldsymbol{b}_{N}\left(i\right)|\boldsymbol{b}_{N}\left(j\right)\right)=1$,
from which we can calculate the corresponding (decimal) indexes $\left(i,j\right)$.
In other words, the algorithm calculates $\mathcal{P}$ efficiently
and in a convenient sparse-matrix notation, since it provides only
the coordinates $\left(i,j\right)$ of the non-zero entries of the
matrix. The matrix $\mathcal{P}$ describes all the possible transitions
between the firing rates, which are not restricted to stationary and
oscillatory solutions only. Eventually, the stationary and oscillatory
solutions of the network may be calculated from $\mathcal{P}$ through
a cycle-finding algorithm (in particular, the stationary solutions
may be considered as oscillations with period $\mathcal{T}=1$). To
the best of our knowledge, the fastest cycle-finding algorithm was
introduced by Johnson \cite{Johnson1975}, and is implemented in the
function \textit{simple\_cycles} of the Python library NetworkX. However,
this approach proved to be even slower than the non-optimized algorithm
of SubSec.~(\ref{sub:Non-Efficient-Algorithms-for-Generating-the-Stationary-and-Oscillatory-Solutions}),
therefore we will not consider it here any further.

\subsection{Examples of Network Topologies \label{sub:Examples-of-Network-Topologies}}

In this section we report two standard examples of network topologies,
whose bifurcation diagrams will be studied in Sec.~(\ref{sec:Results}):
fully-connected networks (SubSec.~(\ref{sub:Fully-Connected-Networks}))
and sparse networks with random synaptic weights (SubSec.~(\ref{sub:Fully-Connected-Networks})).
In these examples, we suppose that the networks are composed of one
excitatory ($E$) and one inhibitory ($I$) population. Two-population
networks are commonly considered a good approximation of a single
neural mass \cite{Grimbert2008}, however our analysis may be extended
to systems composed of an arbitrary number of populations, if desired.
We define $N_{E}$ ($N_{I}$) to be the size of the excitatory (inhibitory)
population, with $N=N_{E}+N_{I}$. Note that $N_{E,I}$ can be arbitrary,
but for illustrative purposes in this section we consider the case
$N_{E}=N_{I}$ (rather than the $N_{E}\backslash N_{I}=4$ ratio experimentally
observed in biological systems \cite{Markram2004}), because we found
that the network complexity increases with the size of the inhibitory
population. Interestingly, the same phenomenon was found to occur
also in multi-population networks with graded activation function
\cite{Fasoli2016}. Moreover, without further loss of generality,
we index the neurons of the excitatory population as $i=0,\ldots,N_{E}-1$
and the inhibitory neurons as $i=N_{E},\ldots,N-1$, so that the synaptic
connectivity matrix $J$ and the stimulus vector $\boldsymbol{I}$
can be written as follows:

\begin{spacing}{0.80000000000000004}
\begin{center}
{\small{}
\[
J=\left[\begin{array}{cc}
\mathfrak{J}_{EE} & \mathfrak{J}_{EI}\\
\mathfrak{J}_{IE} & \mathfrak{J}_{II}
\end{array}\right],\quad\boldsymbol{I}=\left[\begin{array}{c}
\boldsymbol{I}_{E}\\
\boldsymbol{I}_{I}
\end{array}\right].
\]
}
\par\end{center}{\small \par}
\end{spacing}

\noindent $\mathfrak{J}_{\alpha\beta}$, for $\alpha,\beta=E,I$,
is a $N_{\alpha}\times N_{\beta}$ matrix that describes the synaptic
connections from the population $\beta$ to the population $\alpha$.
The entries of the matrices $\mathfrak{J}_{EE}$ and $\mathfrak{J}_{IE}$
must be non-negative, while those of the matrices $\mathfrak{J}_{II}$
and $\mathfrak{J}_{EI}$ must be non-positive, since the populations
$E$ and $I$ are composed of excitatory and inhibitory neurons, respectively.
Moreover, self-connections are not present in biological networks,
so that the main diagonals of the matrices $\mathfrak{J}_{EE}$ and
$\mathfrak{J}_{II}$ should be set to zero (even though, more generally,
our algorithms could be applied also to networks with self-connections).
In a similar way, $\boldsymbol{I}_{\alpha}$ represents the collection
of stimuli to the population $\alpha$. Note that $\mathfrak{J}_{\alpha\beta}$
and $\boldsymbol{I}_{\alpha}$ depend on the structure of the specific
network we study, as we will show below. Finally, we will call $\theta_{\alpha}$
the (homogeneous) firing threshold of all the neurons in the population
$\alpha$.

\subsubsection{Fully-Connected Networks \label{sub:Fully-Connected-Networks}}

Our first example is a fully-connected network with homogeneous intra-population
inputs, whose parameters are:

\begin{spacing}{0.80000000000000004}
\begin{center}
{\small{}
\begin{equation}
\mathfrak{J}_{\alpha\beta}=\begin{cases}
J_{\alpha\alpha}\left(\mathbb{I}_{N_{\alpha}}-\mathrm{Id}_{N_{\alpha}}\right), & \,\mathrm{for}\;\alpha=\beta\\
\\
J_{\alpha\beta}\mathbb{I}_{N_{\alpha},N_{\beta}}, & \,\mathrm{for}\;\alpha\neq\beta
\end{cases},\quad\quad\boldsymbol{I}_{\alpha}=I_{\alpha}\boldsymbol{1}_{N_{\alpha}}.\label{eq:network-structure-example-1}
\end{equation}
}
\par\end{center}{\small \par}
\end{spacing}

\noindent The parameters $J_{\alpha\beta}$ describe the strength
of the homogeneous synaptic connections from the population $\beta$
to the population $\alpha$, while $I_{\alpha}$ represents the stimulus
current to each neuron of the population $\alpha$. Moreover, $\mathbb{I}_{N_{\alpha},N_{\beta}}$
is the $N_{\alpha}\times N_{\beta}$ all-ones matrix (here we use
the simplified notation $\mathbb{I}_{N_{\alpha}}\overset{\mathrm{def}}{=}\mathbb{I}_{N_{\alpha},N_{\alpha}}$),
while $\mathrm{Id}_{N_{\alpha}}$ is the $N_{\alpha}\times N_{\alpha}$
identity matrix and $\boldsymbol{1}_{N_{\alpha}}$ is the $N_{\alpha}\times1$
all-ones vector.

\subsubsection{Sparse Random Networks \label{sub:Sparse-Random-Networks}}

The second example we consider is a sparse network with random synaptic
weights, whose parameters are:

\begin{spacing}{0.80000000000000004}
\begin{center}
{\small{}
\begin{equation}
\mathfrak{J}_{\alpha\beta}=R_{\alpha\beta},\quad\quad\boldsymbol{I}_{\alpha}=\left[\begin{array}{c}
\boldsymbol{0}_{N_{\alpha}-1}\\
I_{\alpha}
\end{array}\right].\label{eq:network-structure-example-2}
\end{equation}
}
\par\end{center}{\small \par}
\end{spacing}

\noindent $R_{\alpha\beta}$ is a $N_{\alpha}\times N_{\beta}$ sparse
random matrix, whose entry $\left[R_{\alpha\beta}\right]_{ij}$, either
for $\alpha=\beta$ and $i\neq j$, or for $\alpha\neq\beta$ and
$\forall i,\:j$, is equal to a non-zero random number with probability
$p_{\alpha\beta}$, while it is equal to zero with probability $1-p_{\alpha\beta}$.
For example, we suppose that the non-zero entries of $R_{\alpha\beta}$
are generated from a homogeneous and uniform (i.e. rectangular) distribution
with support $\left[J_{\alpha\beta}^{\mathrm{min}},J_{\alpha\beta}^{\mathrm{max}}\right]$,
where the parameters $J_{\alpha\beta}^{\mathrm{min}}$ and $J_{\alpha\beta}^{\mathrm{max}}$
describe the minimum and maximum strength of the synaptic connections,
respectively. Moreover, in Eq.~(\ref{eq:network-structure-example-2}),
$I_{\alpha}$ represents the stimulus current to one neuron of the
population $\alpha$, while $\boldsymbol{0}_{N_{\alpha}-1}$ is the
$\left(N_{\alpha}-1\right)\times1$ all-zeros vector (so that the
remaining neurons of the population do not receive any external input).

\section{Results \label{sec:Results}}

\begin{figure}
\begin{centering}
\includegraphics[scale=0.19]{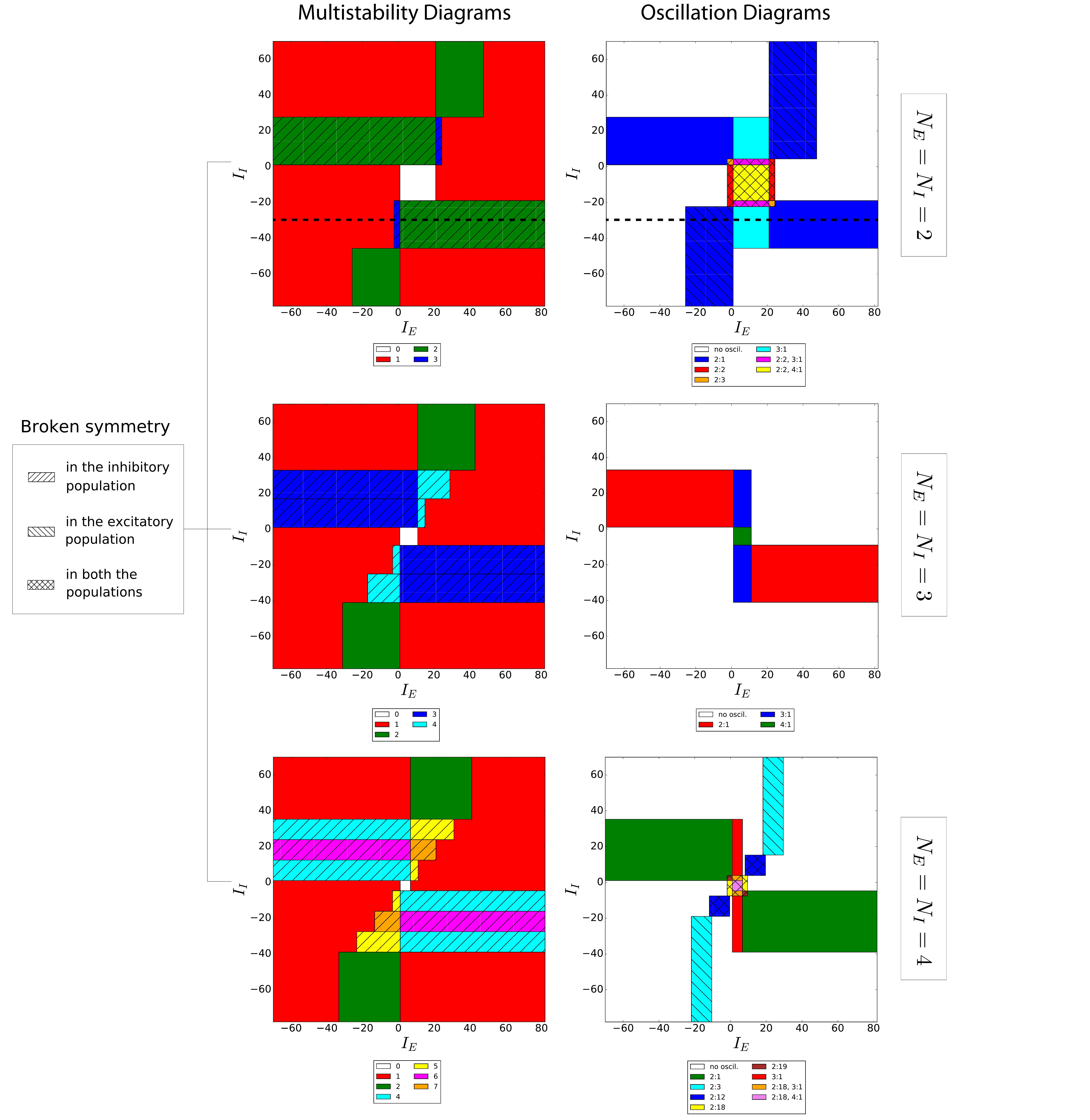}
\par\end{centering}

\protect\caption{\label{fig:example-1} \small\textbf{ Examples of bifurcation diagrams
for fully-connected networks.} This figure is obtained for the network
structure of Eq.~(\ref{eq:network-structure-example-1}), with $N=4$
(top), $6$ (middle), $8$ (bottom), and $J_{EE}=-J_{II}=80$, $J_{IE}=-J_{EI}=70$,
$\theta_{E}=\theta_{I}=1$. The left panels (multistability diagrams)
show how the degree of multistability of the network, namely the number
of stationary solutions, depends on the external currents $I_{E,I}$.
Each color represents a different degree of multistability (e.g.,
blue = tristability). In a similar way, the right panels (oscillation
diagrams) show how the number of oscillatory solutions and their period
are affected by the stimuli. The notation $x$:$y$ reveals the presence
of $y$ different oscillations with period $x$ in a given region
of the diagram. Note that, unlike the multistability diagrams, generally
there is no correspondence among the colors of the three oscillation
diagrams. The script ``Oscillation\_Diagram.py'' assigns the user-defined
colors to each region of the oscillation diagrams regardless of the
specific pairs $x$:$y$ that form that region (e.g., red = $2$:$2$
for $N=4$, while red = $3$:$1$ for $N=8$). Since the total number
of possible regions of the oscillation diagrams is very large and
the actual regions are not known a priori, this prevents the user
from defining too many colors, most of which typically remain unused.}
\end{figure}

In this section we report the bifurcation diagrams generated by our
algorithms for the network topologies described in SubSec.~(\ref{sub:Examples-of-Network-Topologies}).
In Fig.~(\ref{fig:example-1}) we show the (codimension two) bifurcation
diagrams of the fully-connected network (see SubSec.~(\ref{sub:Fully-Connected-Networks}))
in the $I_{E}-I_{I}$ plane, for $N=4,\,6,\,8$. The left and right
panels have been obtained from Eqs.~(\ref{eq:stationary-states-stimulus-range-1})
and (\ref{eq:oscillations-stimulus-range}) respectively, for $\Gamma_{I_{E}}=\left\{ 0,\ldots,N_{E}-1\right\} $
and $\Gamma_{I_{I}}=\left\{ N_{E},\ldots,N-1\right\} $. 
 In particular, we observe that the case $N=6$ is the same considered
in \cite{Fasoli2017}, and whose bifurcation diagram was derived through
hand calculations. Due to the homogeneity of the synaptic connections,
the bifurcation diagrams show symmetric structures, characterized
(also for $N>8$, results not shown) by an increase of the maximum
degree of multistability with the network size and by low-period oscillations.
Interestingly, while spontaneous symmetry-breaking may occur in both
the populations during neural oscillations, only the inhibitory one
may undergo the formation of heterogeneous activity during the stationary
states. As we discuss in SubSec.~(\ref{sub:Future-Directions}),
for any homogeneous multi-population network it is generally possible
to take advantage of this phenomenon, in order to speed up the non-optimized
algorithm of SubSec.~(\ref{sub:Generation-of-the-Set-S-Arbitrary-Networks}).

\begin{figure}
\begin{centering}
\includegraphics[scale=0.2]{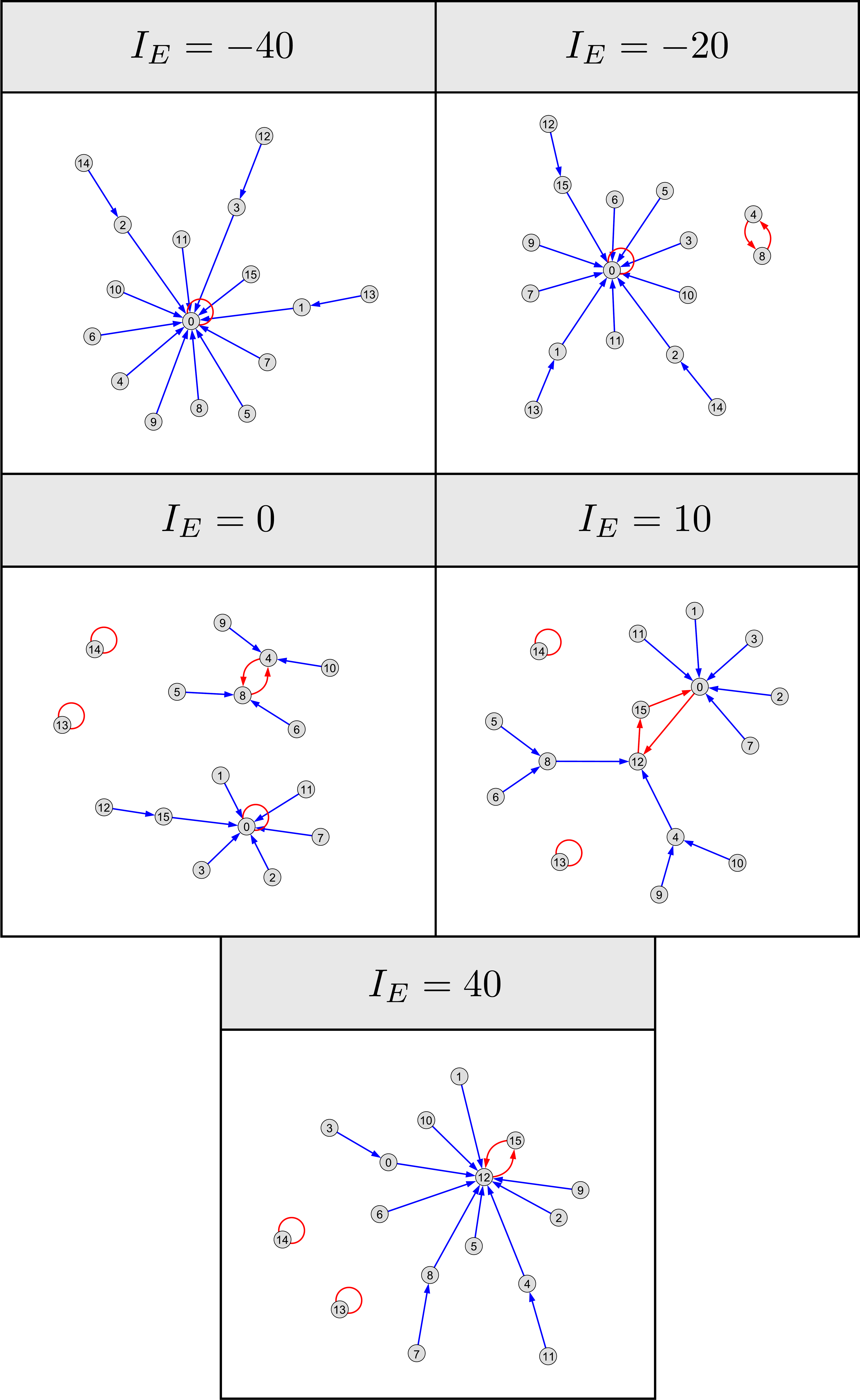}
\par\end{centering}

\protect\caption{\label{fig:example-1-graph} \small\textbf{ Examples of bifurcations
in the state-to-state transition probability matrix.} This figure
shows the changes of dynamics that occur in the fully-connected network
of Eq.~(\ref{eq:network-structure-example-1}) for $N=4$, when varying
the input current $I_{E}$ for $I_{I}=-30$. The nodes in these graphs
are the (decimal representations of the) possible $2^{N}$ firing-rate
states of the network (e.g. the node $3$ corresponds to the firing
rate $\boldsymbol{\nu}=0011$), while the arrows describe the transitions
between the firing rates according to the state-to-state transition
probability matrix (see Eq.~(\ref{eq:conditional-probability})).
The figure highlights in red all the stationary and oscillatory solutions
of the network (compare with the areas crossed by the dashed line
in the top panels of Fig.~(\ref{fig:example-1}), when moving from
left to right).}
\end{figure}

Moreover, in Fig.~(\ref{fig:example-1-graph}) we show some examples
of changes of dynamics that occur in the state-to-state transition
probability matrix of the fully connected network for $N=4$. This
figure shows that the graph of the matrix, as given by Eq.~(\ref{eq:conditional-probability}),
changes its structure as a function of the stimuli. In particular,
the graph undergoes changes in the degree of multistability and in
the period of the oscillations, which are more conveniently described
by Fig.~(\ref{fig:example-1}) (top panels). 

\begin{figure}
\begin{centering}
\includegraphics[scale=0.24]{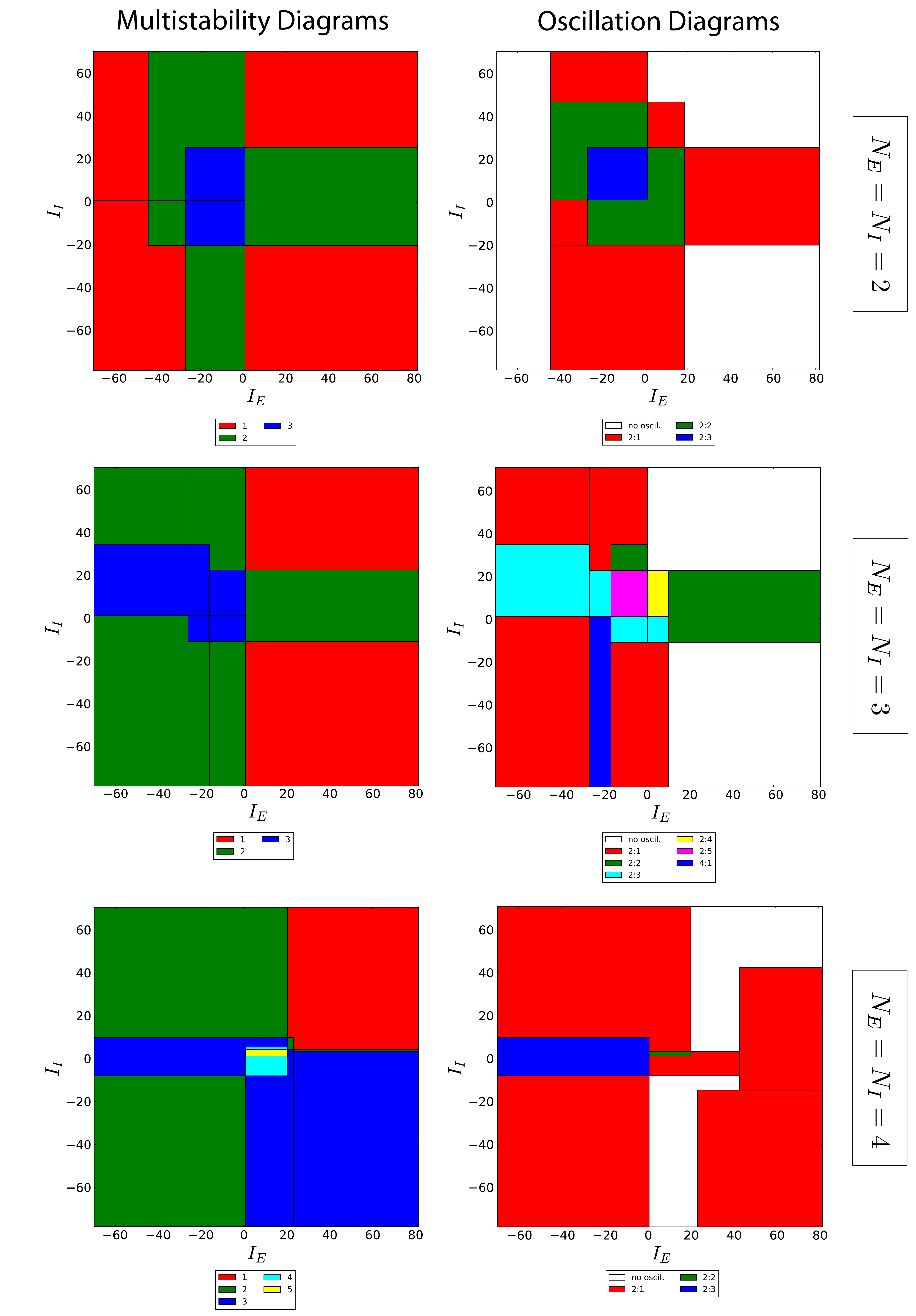}
\par\end{centering}

\protect\caption{\label{fig:example-2} \small\textbf{ Examples of bifurcation diagrams
for sparse random networks.} This figure is obtained for the network
structure of Eq.~(\ref{eq:network-structure-example-2}), with $N=4$
(top), $6$ (middle), $8$ (bottom), the synaptic connectivity matrices
in Tab.~(\ref{tab:Random-connectivity-matrices}), and $\theta_{E}=\theta_{I}=1$.
Similarly to Fig.~(\ref{fig:example-1}), the left (right) panels
show how the stationary (oscillatory) solutions are affected by the
stimuli $I_{E,I}$. For example, in the case $N=8$, both the algorithm
for sparse networks and the non-optimized algorithm derived the multistability
diagram in $0.7$s, since the size of the network is too small for
observing differences between the two algorithms. Moreover, by setting
$\mathcal{T}_{\mathrm{max}}=2,\,3,\,4$, the algorithm for sparse
networks calculated the oscillation diagram in $T_{\mathrm{sparse}}\approx1.2,\,7,\,226$
seconds, respectively (note that in this example $\mathcal{T}_{\mathrm{max}}=2$
is large enough for obtaining the complete diagram, because the network
does not undergo oscillations with period larger than $2$). On the
other hand, the non-optimized algorithm derived the oscillation diagram
in $T_{\mathrm{non-opt}}\approx238$s, through a grid of $5,624$
points in the $I_{E}-I_{I}$ plane (so that the algorithm cannot detect
areas below the resolution $\Delta I=2$). Therefore in this example
the algorithm for sparse networks calculated the oscillation diagram
$T_{\mathrm{non-opt}}\backslash T_{\mathrm{sparse}}\approx198$ times
faster than the non-optimized algorithm.}
\end{figure}

To conclude, Fig.~(\ref{fig:example-2}) shows the bifurcation diagrams
of the sparse random networks (see SubSec.~(\ref{sub:Sparse-Random-Networks}))
in the $I_{E}-I_{I}$ plane, which have been obtained for $N=4,\,6,\,8$
and $\Gamma_{I_{E}}=\left\{ N_{E}-1\right\} ,\,\Gamma_{I_{I}}=\left\{ N-1\right\} $.
As expected, the randomness of the synaptic connections is reflected
in the irregular structure of both the multistability and the oscillation
diagrams, as opposed to the regular structure of the fully-connected
networks shown in Fig.~(\ref{fig:example-1}). In the random networks,
high multistability degrees and high-period oscillations can occur
only by chance, depending on the randomness of the synaptic weights
and on the sparseness of the synaptic topology.

\noindent 

\begin{table}
\begin{centering}
\textbf{\footnotesize{}}%
\begin{tabular}{|c|}
\hline 
\tabularnewline
{\footnotesize{}$J=\left[\begin{array}{cccc}
0 & 80 & -30 & -30\\
91 & 0 & -35 & 0\\
49 & 0 & 0 & -95\\
42 & 0 & -91 & 0
\end{array}\right],\quad J=\left[\begin{array}{cccccc}
0 & 94 & 0 & -30 & -47 & 0\\
97 & 0 & 80 & -44 & 0 & -48\\
0 & 81 & 0 & -30 & -49 & 0\\
0 & 46 & 43 & 0 & -84 & -94\\
0 & 0 & 0 & -89 & 0 & -94\\
0 & 0 & 36 & -100 & -81 & 0
\end{array}\right]$}\tabularnewline
\tabularnewline
{\footnotesize{}$J=\left[\begin{array}{cccccccc}
0 & 89 & 0 & 94 & -46 & 0 & -42 & 0\\
99 & 0 & 100 & 0 & -45 & -40 & 0 & 0\\
0 & 92 & 0 & 0 & 0 & -43 & 0 & 0\\
0 & 0 & 0 & 0 & 0 & -45 & 0 & -39\\
0 & 0 & 0 & 47 & 0 & -91 & -94 & -95\\
0 & 40 & 0 & 0 & -91 & 0 & -80 & -90\\
0 & 0 & 0 & 43 & -90 & -93 & 0 & 0\\
46 & 0 & 0 & 33 & -94 & -89 & -100 & 0
\end{array}\right]$}\tabularnewline
\tabularnewline
\hline 
\end{tabular}
\par\end{centering}{\footnotesize \par}

\protect\caption{\label{tab:Random-connectivity-matrices} \textbf{Synaptic connectivity
matrices used to plot Fig.~(\ref{fig:example-2})}. The non-zero
entries of the matrix $R_{\alpha\beta}$ (see Eq.~(\ref{eq:network-structure-example-2}))
are generated with probabilities $p_{EE}=p_{IE}=0.4$ and $p_{EI}=p_{II}=0.6$,
while the strengths of the synaptic connections are random integers
generated from homogeneous and uniform distributions with $J_{EE}^{\mathrm{min}}=-J_{II}^{\mathrm{max}}=80$,
$J_{EE}^{\mathrm{max}}=-J_{II}^{\mathrm{min}}=100$, $J_{IE}^{\mathrm{min}}=-J_{EI}^{\mathrm{max}}=30$
and $J_{IE}^{\mathrm{max}}=-J_{EI}^{\mathrm{min}}=50$.}
\end{table}

\section{Discussion \label{sec:Discussion}}

We studied how the dynamics of the spin-glass-like neural network
model (\ref{eq:network-equations}) depends on its underlying parameters.
The network is composed of a finite and arbitrary number of neurons
with binary firing rates, that interact through arbitrary (generally
asymmetric) synaptic connections. While in artificial neural networks
with graded (smooth) activation functions this study is performed
by means of standard bifurcation theory \cite{Kuznetsov1998}, the
discontinuity of the activation function of the spin-glass model at
the firing threshold prevents us from using this powerful tool.

Due to the discrete (binary) nature of the network, the number of
possible stationary states and neural oscillations is finite. For
this reason, in SubSec.~(\ref{sub:Non-Efficient-Algorithms-for-Generating-the-Stationary-and-Oscillatory-Solutions})
we developed brute-force algorithms for studying the bifurcations
of the model. By taking advantage of the state-to-state transition
probability matrix, these algorithms find the actual stationary and
oscillatory solutions of the network, and build the corresponding
bifurcation diagram in the parameters space. In particular, in Sec.~(\ref{sec:Results})
we provided examples that show how the network dynamics depends on
the external stimuli, the network size and the topology of the synaptic
connections. Similarly to the case of graded firing-rate network models
\cite{Borisyuk1992,Beer1995,Cessac1995,Pasemann2002,Haschke2005,Fasoli2016},
this analysis revealed a complex bifurcation structure, encompassing
several changes in the degree of multistability of the network, oscillations
with stimulus-dependent frequency, and various forms of spontaneous
symmetry-breaking of the neural activity.

The computational time of these algorithms increases as $2^{N}$ with
the network size, regardless of the network topology. In particular,
the study of neural oscillations proved very challenging due to combinatorial
explosion, causing us to find a compromise between speed and precision.
The best solution we found was to discretize the parameter space of
the oscillation diagram and to run a searching algorithm for every
combination of the parameters on the grid. While this solution allowed
us to perform the study in exponential time, it may generate incomplete
oscillation diagrams, due to the finite grid resolution.

Since biological networks are sparse \cite{Kotter2003}, in SubSec.~(\ref{sub:An-Efficient-Algorithm-for-Generating-the-Stationary-and-Oscillatory-Solutions-in-Sparse-Networks})
we introduced an efficient algorithm that takes advantage of the limited
number of synaptic connections. This algorithm may outperform of several
orders of magnitude the non-optimized approaches introduced in SubSec.~(\ref{sub:Non-Efficient-Algorithms-for-Generating-the-Stationary-and-Oscillatory-Solutions}).
While the computational time of the non-optimized algorithms increases
as $2^{N}$ regardless of the network topology, the speed of the efficient
algorithm increases with the network sparseness and is inversely proportional
to the number of stationary/oscillatory solutions. For these reasons,
the algorithms introduced in SubSec.~(\ref{sub:An-Efficient-Algorithm-for-Generating-the-Stationary-and-Oscillatory-Solutions-in-Sparse-Networks})
prove to be particularly convenient when applied to large sparse networks.
However, for highly dense networks their speed decreases significantly,
therefore in this case the algorithms of SubSec.~(\ref{sub:Non-Efficient-Algorithms-for-Generating-the-Stationary-and-Oscillatory-Solutions})
should be used instead.

While the study of multistability did not reveal any complication,
the study of neural oscillations proved again to be computationally
demanding, also for sparse networks. Our algorithm does not rely on
the discretization of the parameter space, therefore it can detect
areas at any resolution. Nevertheless, it outperforms the non-optimized
algorithm of SubSec.~(\ref{sub:Non-Efficient-Algorithms-for-Generating-the-Stationary-and-Oscillatory-Solutions})
only in detecting oscillations with period $\mathcal{T}\ll Z\overset{\mathrm{def}}{=}\frac{N}{\max_{j}\left(M_{j}-\left|R_{j}\right|\right)}$.
Since $Z$ increases with the network sparseness for a fixed network
size $N$, in sparse networks the algorithm will detect more oscillation
periods in a fixed amount of time, compared to dense networks. While
generally it is not possible to know a priori the effective maximum
period of the oscillations generated by a network (an exception is
represented by symmetric networks, which can sustain only oscillations
with period $\mathcal{T}=2$, see \cite{GolesChacc1985}), in the
sparse random topologies we analyzed, this period was usually small
(see e.g. the right panels of Fig.~(\ref{fig:example-2}), where
we showed that the networks corresponding to the topologies of Tab.~(\ref{tab:Random-connectivity-matrices})
undergo only oscillations with period $2$ and $4$). Oscillations
with period equal to or larger than the network size occur much more
rarely by chance, usually in very small regions of the parameter space
(a formal explanation of this phenomenon is reported in SubSec.~(\ref{sub:New-Insights-into-the-Dynamics-of-Discrete-Network-Models})).
Nevertheless, special (usually very regular) topologies may show large
oscillation periods, therefore a complete analysis of these networks
is beyond the capability of our algorithm for large $N$. Moreover,
while for most of the sparse random topologies we analyzed the algorithm
proved to be very fast, for some networks the script ``Sparse\_Efficient\_Algorithm.py''
generated large sets of candidate oscillatory solutions (only few
of which were actual neural oscillations), resulting in a considerable
slowing down of the bifurcation analysis. This phenomenon occurs in
those networks where the activity of the neurons that receive the
fixed external stimuli does not provide enough information for determining
which oscillations are allowed and which are not. In other words,
depending on the topology of the synaptic connections, in these special
networks the number of oscillatory solutions was strongly influenced
by both the fixed stimuli and the free stimuli that define the parameter
space of the oscillation diagram. Interestingly, this phenomenon occurred
despite the free currents were injected only into a limited number
of neurons by hypothesis (see SubSec.~(\ref{sub:An-Efficient-Algorithm-for-Generating-the-Stationary-and-Oscillatory-Solutions-in-Sparse-Networks})).
In these cases, a grid implementation of the algorithm (similar to
that used in the script ``Non\_Efficient\_Algorithm.py'') could
be considerably faster, since at every point of the grid in the parameter
space the algorithm for sparse networks would generate only actual
oscillatory solutions (see SubSec.~(\ref{sub:An-Efficient-Algorithm-for-Generating-the-Stationary-and-Oscillatory-Solutions-in-Sparse-Networks})).

In the following we discuss the advances of our results with respect
to previous work (SubSec.~(\ref{sub:Progress-with-Respect-to-Previous-Work-on-Bifurcation-Analysis})),
the implications of our work to better understand neural network dynamics
(SubSec.~(\ref{sub:New-Insights-into-the-Dynamics-of-Discrete-Network-Models})),
and future directions that need to be pursued to address the limitations
of our results (SubSec.~(\ref{sub:Future-Directions})).

\subsection{Progress with Respect to Previous Work on Bifurcation Analysis \label{sub:Progress-with-Respect-to-Previous-Work-on-Bifurcation-Analysis}}

Previous studies on the complexity of dynamics in firing-rate network
models focused either on ideal mean-field limits of spin-glass-like
models with discontinuous firing rates (e.g. \cite{Sherrington1976,deAlmeida1978,Mezard1984}),
or on systems with graded (smooth) firing rates (e.g. \cite{Borisyuk1992,Beer1995,Cessac1995,Pasemann2002,Haschke2005,Fasoli2016}).
On the contrary, in this article we studied finite-size networks with
discontinuous activation functions. The analysis of these models is
typically more complex, since we can rely neither on mean-field approximations,
nor on the powerful methods developed in bifurcation theory for smooth
systems. Moreover, our model evolves in discrete-time steps according
to the recurrence relation (\ref{eq:network-equations}). This also
prevents us from using the methods recently developed for the bifurcation
analysis of non-smooth dynamical systems, which can be applied only
to continuous-time models described by non-smooth differential equations
or by differential inclusions \cite{Leine2000,Awrejcewicz2003,Leine2006,Makarenkov2012}.
This proves that the model we considered in this article is particularly
hard to tackle with standard mathematical techniques.

However, in \cite{Fasoli2017} we showed that it is possible to derive
the bifurcation structure of this model by combining a brute-force
search of the stationary and oscillatory solutions of the network,
with the analytical formula of the state-to-state transition probability
matrix. Our proof of concept was limited to the specific case of a
fully-connected network, whose bifurcation structure was calculated
through tedious hand calculations. In the present article we developed
fast algorithms that perform this analysis automatically, for any
topology of the synaptic connections. Therefore our work complements
standard numerical continuation softwares such as the MatCont Matlab
toolbox \cite{Dhooge2003} and XPPAUT \cite{Ermentrout2002}, that
are widely used in neuroscience for the bifurcation analysis of graded
neuronal models (see e.g. \cite{Grimbert2006,Storace2008,Touboul2011,Fasoli2016}).

\subsection{New Insights into the Dynamics of Discrete Network Models \label{sub:New-Insights-into-the-Dynamics-of-Discrete-Network-Models}}

In his pioneering work \cite{Little1974}, Little underlined the importance
of identifying the states that characterize the long-term behavior
of a neural network, which would provide a great simplification in
our comprehension of the network dynamics. In particular, given a
deterministic network model, this problem is equivalent to identifying
the actual stationary and oscillatory solutions for $t\rightarrow\infty$.
While the actual stationary states must be identified out of the set
of $2^{N}$ possible solutions in a network of size $N$, the problem
of selecting the actual oscillations is even more formidable, since
in large asymmetric networks the number of possible oscillatory solutions
is $\sim\left(2^{N}-1\right)!$ (see Appx.~(\hyperref[sec:Appendix-A]{A})).

In our work, we developed algorithms that identify the actual stationary
and oscillatory solutions of Eq.~(\ref{eq:network-equations}). The
algorithms are particularly efficient when applied to sparse networks,
providing great insight into the operation of the network. Moreover,
the algorithms make use of this knowledge to deepen further our understanding
of the network dynamics, by calculating the corresponding bifurcation
structure. Interestingly, our algorithms revealed the formation of
complex bifurcation diagrams also in small networks. The diagrams
strongly depend on the network size, $N$, and on the topology of
the synaptic connections, $J$. However, a detailed analysis of the
network dynamics as a function of $N$ and $J$ is computationally
very expensive (due to the large number of possible topologies for
each fixed network size) and beyond the purpose of this article. Nevertheless,
our study revealed that the spin-glass-like network model (\ref{eq:network-equations})
can undergo three different kinds of bifurcations.

The first is a change in the degree of multistability of the network.
This phenomenon occurs through the formation and destruction of stationary
solutions, similarly to the \textit{limit-point} (also known as \textit{saddle-node})
bifurcations that occur in graded models \cite{Kuznetsov1998}.

The second kind of bifurcation is the formation of neural oscillations.
If this phenomenon occurs through the annihilation of a stationary
state (see e.g. the formation of the period-$4$ oscillation in the
middle panels of Fig.~(\ref{fig:example-1})), then it may be interpreted
as the discrete counterpart of the \textit{Andronov-Hopf} bifurcation
of graded models \cite{Kuznetsov1998}. Otherwise, the number of oscillatory
solutions may change with no variation in the number of stationary
states (see e.g. the transition between the blue and red areas in
the top-right panel of Fig.~(\ref{fig:example-1})), similarly to
the \textit{limit point of cycles} bifurcation of graded models \cite{Kuznetsov1998}.
It is interesting to observe that usually large-period oscillations
occur more rarely and in smaller regions of the parameter space (for
example, compare the green area of the period-$4$ oscillation in
the middle-right panel of Fig.~(\ref{fig:example-1}), with the blue
and red areas of the period-$3$ and period-$2$ oscillations). The
reason of this phenomenon can be easily deduced from Eq.~(\ref{eq:oscillations-stimulus-range}):
the larger the period of the oscillation, the more probable is that
the function $\Phi_{I}=\underset{t\in\mathscr{T}}{\max}\left(\cdot\right)$
(respectively, $\Psi_{I}=\underset{t\in\mathscr{T}}{\min}\left(\cdot\right)$)
will be large (respectively, small). In turn, this implies that the
range $\left(\Phi_{I},\Psi_{I}\right]$ where the large-period oscillation
occurs will be narrower, and in particular if $\Phi_{I}\geq\Psi_{I}$
the oscillation will not occur at all.

The third kind of bifurcation is characterized by spontaneous symmetry-breaking.
This phenomenon is similar to the \textit{branching-point} bifurcations
\cite{Kuznetsov1998} that occur in homogeneous multi-population graded
models (see e.g. \cite{Fasoli2016}). Here we observe the formation
of heterogeneous intra-population neural activity, even if Eq.~(\ref{eq:network-equations})
does not contain any term that breaks explicitly the network symmetry.
In particular, the symmetry may be broken at the stationary states
or during oscillations, in both excitatory and inhibitory neural populations,
depending on the network topology and parameters.

To conclude, we observe that our algorithms also calculate the number
of stationary and oscillatory solutions, as well as the number of
areas with distinct degrees of multistability, the number of oscillations
with a given period, etc. We propose that these measures could be
used to quantify rigorously the complexity of the neural dynamics
in relation to the network size.

\subsection{Future Directions \label{sub:Future-Directions}}

Neural networks with discrete activation functions may be considered
as caricatures of the corresponding graded models. By discretizing
the activation function, we gain the possibility to solve exactly
the network equations \cite{Fasoli2017}, but currently it is not
known to which extent this approximation oversimplifies the dynamical
behavior of the system. For this reason, in future work we will investigate
a potential reduction of the underlying neural complexity of the network
by comparing the two classes of models.

Moreover, note that while in this article we introduced new algorithms
for the bifurcation analysis of Eq.~(\ref{eq:network-equations}),
we did not perform an exhaustive analysis of the relationship between
the neural dynamics and the network parameters. In Sec.~(\ref{sec:Results})
we showed some relevant examples of the codimension two bifurcation
diagrams for specific networks sizes and topologies, while in future
work we will systematically investigate, through a large scale analysis,
to which extent the network parameters correlate with the neural complexity.

To conclude, we observe that while we introduced fast algorithms for
the bifurcation analysis of sparse networks, efficient solutions for
dense networks are still missing. Ad hoc solutions can be developed
for specific network topologies. For example, in the case of multi-population
fully-connected networks with homogeneous weights, such as that considered
in Eq.~(\ref{eq:network-structure-example-1}), it is possible to
prove that the symmetry of the excitatory populations is never broken
when the network is in a stationary state. The same phenomenon occurs
in homogeneous multi-population networks with graded activation function
\cite{Fasoli2016}. Accordingly, the method described in SubSec.~(\ref{sub:Generation-of-the-Set-S-Arbitrary-Networks})
can be easily modified in order to include in the set $\mathscr{S}$
only the states with homogeneous excitatory firing rates. In the case
of the two-populations network described by Eq.~(\ref{eq:network-structure-example-1}),
this approach would generate a set with cardinality $\left|\mathscr{S}\right|=2^{1+N_{I}}$,
rather than $\left|\mathscr{S}\right|=2^{N_{E}+N_{I}}$, reducing
considerably the computational time for $N_{E}\gg1$. However, this
idea can be applied only to specific network topologies, and a general
method for speeding up the bifurcation analysis of dense networks
with arbitrary topology is still missing. This represents another
challenge that needs to be addressed in future work, and that will
complete the tools at our disposal for studying the dynamical behavior
of binary neuronal network models.

\section*{Appendix A: Asymptotic Estimation of the Number of Possible Oscillatory
Solutions in Large Networks \label{sec:Appendix-A}}

In this appendix we want to determine how the number of possible oscillatory
solutions in an asymmetric network of size $N$ grows for $N\rightarrow\infty$.
Given any finite $N$, we define $\mathcal{N}\overset{\mathrm{def}}{=}2^{N}$
and we observe that $\binom{\mathcal{N}}{k}$ is the total number
of sets (known as $k$-combinations) containing $k$ distinct neural
states among which the network may oscillate. For example, for $k=3$,
the oscillations $0\rightarrow1\rightarrow2\rightarrow0$ and $0\rightarrow1\rightarrow3\rightarrow0$
occur among the states in the $3$-combinations $\left\{ 0,1,2\right\} $
and $\left\{ 0,1,3\right\} $, respectively. Now we observe that any
given set of $k$ states gives rise to $\left(k-1\right)!$ distinct
oscillations. For example, $0\rightarrow1\rightarrow2\rightarrow0$
and $0\rightarrow2\rightarrow1\rightarrow0$ are two distinct oscillations
generated by the set $\left\{ 0,1,2\right\} $, while the oscillations
$0\rightarrow1\rightarrow2\rightarrow0$ and $1\rightarrow2\rightarrow0\rightarrow1$
are (circularly) identical. Therefore, by summing over all the oscillation
periods $k$ from $2$ to $\mathcal{N}$ (the case $k=1$ corresponds
to the stationary states and therefore is not considered), we get
that the total number of distinct oscillations is $\sum_{k=2}^{\mathcal{N}}\binom{\mathcal{N}}{k}\left(k-1\right)!$
(see also \cite{Johnson1975}). Then, by observing that:

\begin{spacing}{0.80000000000000004}
\begin{center}
{\small{}
\begin{align*}
\left(k-1\right)!= & \int_{0}^{\infty}x^{k-1}e^{-x}\,dx\\
\\
\sum_{k=0}^{\mathcal{N}}\binom{\mathcal{N}}{k}x^{k}= & (1+x)^{\mathcal{N}},
\end{align*}
}
\par\end{center}{\small \par}
\end{spacing}

\noindent we finally get the following asymptotic expansion:

\begin{spacing}{0.80000000000000004}
\begin{center}
{\small{}
\begin{align*}
 & \sum_{k=2}^{\mathcal{N}}\binom{\mathcal{N}}{k}\left(k-1\right)!=\sum_{k=2}^{\mathcal{N}}\binom{\mathcal{N}}{k}\int_{0}^{\infty}x^{k-1}e^{-x}\,dx=\int_{0}^{\infty}\frac{\left(1+x\right){}^{\mathcal{N}}-\mathcal{N}x-1}{x}e^{-x}\,dx\\
\\
 & \sim\int_{0}^{\infty}x^{\mathcal{N}-1}e^{-x}\,dx=\left(\mathcal{N}-1\right)!
\end{align*}
}
\par\end{center}{\small \par}
\end{spacing}

\noindent in the limit $\mathcal{N}\rightarrow\infty.\;\;\blacksquare$

\section*{Acknowledgments}

We thank Anna Cattani for useful feedback. This research was supported
by the Autonomous Province of Trento, Call \textquotedblleft Grandi
Progetti 2012,\textquotedblright{} project \textquotedblleft Characterizing
and improving brain mechanisms of attention - ATTEND\textquotedblright .

\noindent The funders had no role in study design, data collection
and analysis, decision to publish, interpretation of results, or preparation
of the manuscript.

\bibliographystyle{plain}
\bibliography{Bibliography}

\end{document}